\documentclass[preprint2,twocolumn,times,tighten]{aastex63}

\usepackage{graphicx,times}
\usepackage{subfigure}
\newcommand{\be}{\begin{equation}}
\usepackage{threeparttable}
\usepackage{booktabs}
\newcommand{\ee}{\end{equation}}
\newcommand{\bea}{\begin{eqnarray}}
\newcommand{\eea}{\end{eqnarray}}
\usepackage{color}
\newcommand*{\al}{\color{black}}
\newcommand*{\xu}{\color{black}}

\usepackage{enumerate}
\usepackage{amsmath}
\usepackage{cases}
\usepackage{longtable}
\usepackage{hyperref}
\usepackage{epstopdf}
\usepackage{amsmath,bm}
\usepackage{amssymb}
\usepackage{natbib}
\usepackage{morefloats}
\usepackage{multirow}
\usepackage{array}
\usepackage{verbatim}
\usepackage{lineno}

\shorttitle{Mirror diffusion of cosmic rays}
\shortauthors{Lazarian \& Xu}

\begin{document}

\title{Diffusion of cosmic rays in MHD turbulence with magnetic mirrors}

\email{lazarian@astro.wisc.edu; sxu@ias.edu}

\author{A. Lazarian}
\affiliation{Department of Astronomy, University of Wisconsin, 475 North Charter Street, Madison, WI 53706, USA}
\affiliation{Center for Computation Astrophysics, Flatiron Institute, 162 5th Ave, New York, NY 10010}
\author{Siyao Xu}
\affiliation{Institute for Advanced Study, 1 Einstein Drive, Princeton, NJ 08540, USA\footnote{Hubble Fellow}}

\begin{abstract}

As the fundamental physical process with many astrophysical implications, 
the diffusion of cosmic rays (CRs) is determined by their interaction with 
magnetohydrodynamic (MHD) turbulence.
We consider the magnetic mirroring effect
arising from MHD turbulence on the diffusion of CRs. 
Due to the intrinsic superdiffusion of turbulent magnetic fields, CRs with large pitch angles that undergo mirror reflection, i.e., bouncing CRs, are not trapped between magnetic mirrors, but move diffusively along 
the {\xu turbulent} magnetic field, 
leading to a new type of parallel {\xu diffusion, i.e., mirror} diffusion.
This {\xu mirror} diffusion 
is in general slower than the diffusion of non-bouncing CRs with small pitch angles 
that undergo gyroresonant scattering. 
The critical pitch angle at the balance between magnetic mirroring and pitch-angle scattering is important for determining the diffusion coefficients of both bouncing and non-bouncing CRs
and their scalings with the CR energy. 
We find non-universal energy scalings of diffusion coefficients, depending on the properties of MHD turbulence. 

\end{abstract}

\section{Introduction}

Charged energetic particles or cosmic rays (CRs) are an important ingredient in the physical processes in space and astrophysical environments. 
It is customary to use the term 
``energetic particles" in space physics. 
The theoretical understanding on 
their acceleration and diffusion in the 
Solar atmosphere, solar wind, Earth magnetosphere, and heliosphere
is important for studying the 
properties of the interplanetary magnetic field, 
solar modulation of Galactic CRs, 
and space weather forecasting 
\citep{Par65,Jo71,SinH01}.

The energetic particles with higher energies outside our direct neighborhood, i.e., of Galactic and extragalactic origin,
are usually referred to as CRs. 
The knowledge on 
the acceleration and diffusion of CRs 
is essential for 
probing their sources, 
explaining their chemical composition, 
studying their roles in 
ionizing molecular gas and circumstellar discs
(e.g., \citealt{Schlk16,Padd18}),
driving galactic winds 
(e.g., \citealt{Ipa75,Hol19}),
and feedback heating in clusters of galaxies
(e.g., \citealt{Guo08,Brun14}),
as well as modeling the synchrotron foreground emission for 
cosmic microwave
background (CMB) radiation
and redshifted $21$ cm radiation
(e.g., \citealt{Chf02,Chh12}). 
In this work, we focus on the diffusion physics that is generally applicable to energetic particles of Solar origin and CRs. 
Thus we do not distinguish between them and will only use the term ``CRs".


The diffusion of CRs is governed by their interaction with turbulent magnetic fields, {\al as well as the magnetic fluctuations arising from the instabilities induced by CRs, i.e.,}  {\xu the streaming instability 
\citep{Kulsrud_Pearce}. 
In this work we focus on 
the former type of interaction.}
{\al The CRs interaction with magnetic turbulence} has been a subject of intensive research for decades
\citep{Jokipii1966,Kulsrud_Pearce,SchlickeiserMiller,Giacalone_Jok1999}. {\xu However, ad hoc models for magnetohydrodynamic (MHD) turbulence were adopted in early studies.} 
\footnote{\xu The model of isotropic MHD turbulence
(see \citealt{Schlickeiser02})
and 2D + slab superposition model of solar wind turbulence
\citep{Mat90}
adopted in early studies 
are in contradiction with
MHD turbulence simulations 
and are also challenged by solar wind observations 
(e.g., \citealt{Hor08}).}

{\xu The recently developed 
theories of MHD turbulence 
\citep{GS95,LV99,LG01} have been numerically tested}
(\citealt{CV00,MG01,CLV_incomp,CL03,Bere14,KowL10,Kow17}, see also the book by \citealt{BL19}), 
which are also supported by solar wind observations  
(e.g., \citealt{Hor08,Luo10,For11}).
{\xu The development of MHD turbulence theories} 
allows important advances 
{\xu in studying the}
pitch-angle scattering, stochastic acceleration, 
spatial diffusion of CRs along the magnetic field
\citep{Chan00,YL02,YL04,Brunetti_Laz,YL08,Ly12,XY13,XLb18,Sio20,Lem20},
superdiffusion and 
diffusion perpendicular to the mean magnetic field 
\citep{YL08,LY14},
propagation of CRs in weakly ionized astrophysical media 
\citep{Xuc16},
and interactions of CRs with relativistic MHD turbulence 
\citep{Demi20}. 
These studies bring significant 
changes to  
{\xu the standard diffusion models of CRs}
based on the ad hoc models of MHD turbulence 
(e.g., \citealt{Mat90,Kota_Jok2000,Qin2002})
and shed light on some long-standing problems and observational puzzles
(e.g., \citealt{Palmer:1982,Evol14,Lop16,Krum20}).
{\xu Naturally, 
the propagation of CRs should be modeled using the tested MHD turbulence theories 
in order to explain multifrequency observations and direct CR measurements  (e.g.,
\citealt{Nav13,Coh16,Orl18,Gabi19,Amat21,Forn21}).
{We are still far from fully understanding the diffusion of CRs to interpret the observations near the Earth and 
in the vicinity of CR sources and their differences. 
This motivates us to further study the fundamental physics of CR diffusion in this work. }



{\xu For CRs interacting with magnetic irregularities,}
in addition to the pitch-angle scattering, 
{\xu it has long been known that the CR propagation} can also be affected by magnetic mirror reflection 
\citep{Fer49,Noe68,CesK73,Klep95,Chanmc00}. 
{\xu The magnetic mirroring effect} was explored, for instance, {\al for solving} the $90^\circ$ problem of the 
quasilinear theory (QLT,
\citealt{Jokipii1966})
for pitch-angle scattering. 
In these early studies, the mirroring effect was invoked for trapping CRs that bounce back and forth between two mirror points,
but 
it has not been considered 
{\xu in the context of the numerically tested} 
modern theories of MHD turbulence. 

In MHD turbulence, compressions of magnetic fields, which are generated by pseudo-Alfv\'{e}n modes in 
an incompressible medium and slow and fast modes in a compressible medium, 
naturally give rise to the mirroring effect
over a range of length scales following the energy cascade of turbulence. 
The interaction of CRs with magnetic mirrors is regulated by the dynamics of turbulent magnetic fields. 
{\xu In particular, 
this work will demonstrate that} the intrinsic perpendicular superdiffusion of turbulent magnetic fields
\citep{LV99,Eyin13,LY14}
and its interaction with the parallel diffusion
should be taken into account when 
studying the CR diffusion subject to the mirroring effect.

By using the numerically tested scalings of MHD turbulence 
\citep{CL03},
\citet{XL20} (hereafter XL20)
investigated the scattering of CRs with the mirroring effect included.  
There we confirmed the dominant role of fast modes in gyroresonant scattering, 
and we
found that the resulting 
diffusion coefficient can be significantly smaller than that in the absence of magnetic mirroring. 

{\xu In this work, 
by taking into account the intrinsic dynamics of MHD turbulence, i.e., magnetic field perpendicular superdiffusion, 
we investigate the effect of magnetic mirroring on the parallel diffusion of CRs. 
This new diffusion mechanism arising from the mirroring effect in MHD turbulence should be considered together with other diffusion processes related to scattering and streaming of CRs for a more comprehensive description of CR propagation.
Here we focus on the fundamental physics of the mirror diffusion mechanism and the formulation of its diffusion coefficient. 
Its applicability to various astrophysical media and 
its confrontation with observations will be studied in our future work 
(see e.g., \citealt{Xu21}).}

In what follows, in Section 2, 
we introduce the perpendicular superdiffusion of both turbulent magnetic fields and CRs 
and its effect on the parallel diffusion of bouncing CRs. 
In Sections 3 and 4, we formulate the diffusion coefficients of bouncing CRs in compressible and incompressible MHD turbulence, respectively. 
In Section 5, we consider the exchange between bouncing and non-bouncing CRs and 
discuss the averaged diffusion coefficient
{\xu on scales much larger than their mean free paths}. 
Finally, 
the discussion and the summary of our main results are given in Sections 6 and 7.

\section{Spatial diffusion of bouncing particles }

{\xu The mirroring effect in MHD turbulence causes diffusion of CRs along turbulent magnetic fields. 
This diffusion parallel to the {\it local} magnetic field,
i.e., "parallel diffusion", is affected by the superdiffusion of turbulent magnetic fields in the direction perpendicular to the mean magnetic field. 
Below we will discuss in detail these two types of diffusion and their relation.}


\subsection{Magnetic mirrors in MHD turbulence}
\label{ssec: magbet}

MHD turbulence can be decomposed into Alfv\'{e}n, slow, and fast modes
\citep{GS95,LG01,CL02_PRL,CL03}. 
Alfv\'{e}n modes induce magnetic field wandering, 
which accounts for the superdiffusion of CRs perpendicular to the mean magnetic field \citep{LV99,Eyin13,LY14}. 
Slow modes are passively mixed by Alfv\'{e}n modes and have the same anisotropic scaling as 
Alfv\'{e}n modes \citep{LG01,CL03}. 
Fast modes have an independent energy cascade and isotropic energy distribution \citep{CL02_PRL, CL03, KowL09}.
{\al Magnetic compressions arising from} slow and fast modes can act as magnetic mirrors that result in 
bouncing of particle among the mirror points. 
Their detailed statistical properties are presented in Appendix.

The magnetic mirroring effect caused by static magnetic bottles
is well known in plasma physics 
(e.g., \citealt{Post58,Bud59,Noe68,Kulsrud_Pearce}).
A particle with the Larmor radius smaller than the variation scale of the magnetic field
preserves its adiabatic invariant, i.e. $p_{\bot}^2/B=$const. 
It implies 
\begin{equation}
\frac{p_\bot^2}{B_0}=\frac{p^2}{B_0+\delta b}
\end{equation}
for the condition of magnetic mirroring, 
where $B_0$ and $B_0 + \delta b$ are the 
magnetic field strengths in the weak and strong magnetic field regions, 
and $p$ is the total momentum of the particle. 
{\al As the total momentum is preserved, it is easy to see that} the particles with $\mu < \mu_{lc}$, where $\mu$ is the pitch-angle cosine and $\mu_{lc}$ satisfies 
\begin{equation}\label{mu}
\mu_{lc}^2=\cos^2 \theta_{lc}=\frac{\delta b}{B_0+\delta b},
\end{equation}
are subject to magnetic mirroring. 
Here $\theta_{lc}$ is angular size of the loss cone for escaping particles with smaller pitch angles. 
For slow and fast modes in MHD turbulence with a spectrum of magnetic fluctuations, 
there is $\delta b=b_k$.
When $b_k$ at wavenumber $k$ is significantly smaller than the mean magnetic field strength $B_0$, 
the above expression can be approximated by 
\begin{equation}\label{mubk}
\mu_{lc}^2\approx \frac{b_k}{B_0}.
\end{equation}

The amplitude of $b_k$ depends on the driven 
magnetic perturbations $\delta B_s$ and $\delta B_f$ of slow and fast modes. 
Their relation depends on the compressibility of the medium and plasma $\beta$, where $\beta = P_\text{gas}/P_\text{mag} > 1$, , and $P_\text{gas}$ and $P_\text{mag}$ are gas/plasma and magnetic pressures.
In a in high-$\beta$ medium, with the sonic Mach number $M_s = V_L / c_s < 1$, 
 where $V_L$ is the driven turbulent velocity,
$c_s$ is the sound speed, 
$\delta B_s$ is expected to be larger than
$\delta B_f$. 
In the opposite limit of magnetically dominated medium with $\beta<1$, 
there is $\delta B_s < \delta B_f$.

In the literature, 
particles with $\mu<\mu_{lc}$ are considered ``trapped"
in magnetic bottles and thus they cannot diffuse. 
However, 
as we will show below (Section \ref{sssec: pardiff}), this is an erroneous notion. 
In fact, for 3D propagation in MHD turbulence,
magnetic mirroring results 
in a new type of parallel diffusion of bouncing particles.


\subsection{Perpendicular superdiffusion}
\label{sec: bounc}

CRs following magnetic field lines
\footnote{In the presence of turbulence, CRs follow the magnetic flux tubes averaged within their gyro-orbit. 
We nevertheless adopt the generally accepted way to describe magnetic fields as ``magnetic field lines".}
undergo pitch-angle scattering via the interaction with small-scale magnetic fluctuations 
and bouncing among magnetic mirrors (see above).
In the meantime, 
the dispersion of their trajectories in the direction perpendicular to the mean magnetic field is determined by 
the dispersion of magnetic field lines. 
This dispersion increases with the propagation distance of CRs along the magnetic field 
and accounts for the perpendicular diffusion of CRs, as pointed out by 
\citet{Jokipii1966}.
The superdiffusive nature of this dispersion {within the range of length scales of strong MHD turbulence} was later found by 
\citet{LV99}
when they analytically quantified the magnetic field wandering induced by the Alfv\'{e}nic component\footnote{The separation of MHD turbulence into Alfvenic, slow and fast modes was demonstrated numerically in Cho \& Lazarian (2002, 2003). The dominance of Alfvenic modes in terms of inducing magnetic field wandering was shown analytically in \citet{LV99}.} of MHD turbulence for 
both super-Alfv\'{e}nic ($M_A = V_L/V_A>1$) and sub-Alfv\'{e}nic ($M_A <1$) turbulence, 
where $V_A$ is the Alfv\'{e}n speed. 
We note that the perpendicular superdiffusion is with respect to the mean magnetic field in 
sub- and trans-Alfv\'{e}nic ($M_A=1$) turbulence, and 
with respect to the local mean field, i.e., the magnetic field averaged over 
scales less than $l_A=L M_A^{-3}$ in super-Alfv\'{e}nic turbulence
\citep{Lazarian06}, where $L$ is the turbulence driving scale, 
and $l_A$ is the scale where the local turbulent velocity becomes equal to $V_A$. 
For the turbulent motions on scales larger than $l_A$, the effect of magnetic field is subdominant, and they gradually get isotropic, similar to the hydrodynamic Kolmogorov turbulence. 

For sub-Alfv\'{e}nic turbulence, there exists a scale $l_\text{tran}=LM_A^2$ for the transition from the weak turbulence over the scales $[l_\text{tran}, L]$ to the strong turbulence on smaller length scales. 
In the weak turbulence regime, 
the magnetic fields exhibit normal diffusion, i.e. the mean squared separation between magnetic field lines
{\xu increases with $s$, where $s$ is the distance measured along the magnetic field lines \citep{LV99}.}

We next consider 
the divergence of magnetic field lines over scales less than $l_\text{tran}$. 
Naturally, 
when one follows the magnetic field line over the parallel scale $l_\|$ of a turbulent eddy, 
the magnetic field line has its perpendicular displacement equal to the transverse size $l_{\bot}$ of the eddy. 
As $l_\bot$ can be both positive and negative, 
the dispersion $\langle y^2 \rangle$ of magnetic field lines increases as 
\begin{equation}\label{y2}
d\langle y^2 \rangle\approx l_\bot^2 \frac{ds}{l_\|},
\end{equation}
where $ds$ is the distance measured along the magnetic field line, 
and the bracket denotes an ensemble average. 
Using the above relation and the scaling relation between $l_{\|}$ and $l_{\bot}$ for 
sub-Alfv\'{e}nic turbulence 
(\cite{LV99}, see also Appendix \ref{sec:basic}), one can find 
\begin{equation}\label{y3}
d\langle y^2 \rangle\approx \langle y^2 \rangle^{2/3} M_A^{4/3} L^{-1/3} ds.
\end{equation}
If the expression on the right-hand side were 
``constant$\times ds$", Eq. \eqref{y3} would describe the random walk of field lines. 
However, it has a power-law dependence on the field line separation with a positive power index, 
i.e. $\propto \langle y^2 \rangle^{2/3}$,
leading to an accelerating superdiffusion as magnetic field lines spread out. 
The numerical demonstration of the superdiffusion of turbulent magnetic fields can be found in, e.g., 
\citet{Laz04,Bere13}.

The physical explanation of the effect is the following. 
As we follow the magnetic field line the distance $\sim s$, 
the divergence rate of field lines  increases with larger and larger turbulent eddies contributing to the dispersion of field line separations.
A {\xu similar} effect in hydrodynamic turbulence is related to the accelerated separation of a pair of test particles with time, which is known as Richardson dispersion
\citep{Rich26}.\footnote{\al In the presence of magnetic field, this results in the Richardson dispersion of magnetic field lines with time 
\citep{Eyink2011,Eyin13}, 
{\xu as a time-dependent analog of the superdiffusion of magnetic field lines in space.}}
The accelerated divergence of magnetic field lines results in the superdiffusive field line separations. 
The analogy between the Richardson dispersion and the superdiffusion of magnetic fields 
introduced in \citet{LV99} was discussed in detail in \citet{Eyink2011}.

Eq. \eqref{y3} also applies to the dispersion of separations of CRs that 
follow magnetic field lines.
The resulting superdiffusive perpendicular divergence of CR trajectories was  studied in
\citet{YL08,LY14} {\al in two regimes, one describing the ballistic motion of CRs along the magnetic field, the other describing the diffusion of CRs along the magnetic field.} Both regimes were numerically tested in 
\citet{XY13}.
{\al For the ballistic propagation, the corresponding dispersion} of separations of particles in the perpendicular direction is
\begin{equation}
\langle y^2\rangle \sim l_\bot^2 \approx \frac{s^3}{L} M_A^4, ~~~M_A<1, ~s<\lambda_\|,
~ s< L.
\label{perpen1}
\end{equation}
Here $s$ is the distance of particles measured along the magnetic field line, 
{\xu and $\lambda_\|$ is the CR parallel mean free path.} 
$y$ can be identified with the displacement of particles perpendicular to the mean magnetic field. 
{\xu Note that the dependence $\langle y^2\rangle\propto s^3$ in the above expression is also applicable to 
super-Alfv\'{e}nic turbulence at scales less than $l_A$, i.e., $s<l_A$.}

In the presence of the parallel diffusion, i.e. over the distance $ s > \lambda_\|$, the perpendicular dispersion of CRs is modified to scale as $\langle y^2\rangle \propto s^{3/2}$.  Naturally, this is still in a superdiffusion regime
that arises from the superdiffusive magnetic field line dispersion.
Therefore, {in strong MHD turbulence, i.e., on scales smaller than $l_A$ in super-Alfv\'{e}nic turbulence and $l_\text{tran}$ in sub-Alfv\'{e}nic turbulence, 
for both ballistic and diffusive CR propagation} 
{\xu along the turbulent magnetic field, 
CR trajectories get 
separated superdiffusively in the direction perpendicular to the mean magnetic field, 
although the law of superdiffusion differs in the two cases.} 
A detailed discussion on the perpendicular (super)diffusion of CRs in different turbulence regimes is provided in 
\citet{LY14} ({\al see Table 1 in \citealt{Lsup19}}).

{We note that the concept of perpendicular superdiffusion of CRs contradicts to some existing views of CR transport, e.g. the Non-Linear Guiding Center Theory (NLGCT) 
\citep{Matt03}
that is formulated using the 2D/slab model of MHD turbulence. 
As the (super)diffusion behavior of CRs strongly depends on the properties of MHD turbulence, 
CR (super)diffusion should be studied using the tested model of MHD turbulence, instead of synthetic models. }

\subsection{\xu Parallel mirror diffusion}
\label{sssec: pardiff}

The effect of magnetic mirrors on trapping particles was extensively studied in the literature (e.g., \citealt{Noe68,Kulsrud_Pearce,CesK73}). 
There it was assumed that the trapped particles undergo oscillatory motions between two mirror points of a magnetic bottle, without cumulative diffusion along the magnetic field. 
As the main difference between this work and earlier studies, 
here we consider the existence of Alfv\'{e}nic component of turbulence and the resulting perpendicular superdiffusion of magnetic fields and CRs. 
{\al As we explain below, } in the presence of Alfv\'{e}nic turbulence, 
the superdiffusion of magnetic fields does not allow the particles 
to be trapped in the same magnetic bottle.

For every crossing of a magnetic bottle induced by the compressive component of MHD turbulence with a size $s$, 
the particle experiences perpendicular superdiffusion with the perpendicular displacement $y$ given 
{\xu in Section \ref{sec: bounc}.}
As a result, the particle 
escapes the magnetic bottle within one crossing time and then encounters another magnetic bottle within a perpendicular distance $y$ from the previous magnetic bottle. 
As a result, the bouncing with the mirror points of different magnetic bottles leads to the diffusive motion of particles along  
{\xu turbulent magnetic field lines.} 
We term this 
diffusion {\xu parallel to local magnetic field
arising from the mirroring effect as
``parallel mirror diffusion"} to distinguish it from the traditional CR parallel diffusion associated with the resonance scattering 
(see \citealt{Schlickeiser02}).  
We illustrate the parallel {\xu mirror} diffusion in Fig. \ref{fig: sket}.
CRs that follow diffusing magnetic field lines are not trapped, but move diffusively parallel to the turbulent magnetic field when bouncing with different magnetic mirrors. 
{\xu Through this paper, ``bouncing" and ``mirroring" are equivalent.}

{\al Due to the 3D character of CR motions and the complexity of magnetic field structure at different $M_A$,}
{\xu it is important to clarify that the parallel diffusion of CRs in MHD turbulence}
{\al is the diffusion with respect to the local magnetic field sampled by CRs. 
At $M_A<1$,} 
{\xu the local magnetic field has the direction close to that of the global mean magnetic field, 
while at $M_A>1$, 
the magnetic field direction changes significantly 
on scales larger than $l_A$. 
In the latter case, we consider the parallel mirror diffusion on scales smaller than $l_A$. 
The parallel mirror diffusion of CRs
is accompanied by the dispersion of their trajectories in the direction perpendicular to the mean magnetic field 
(see Fig. \ref{fig: sket}). 
The perpendicular superdiffusion of CRs can be observed when their initial separations in space are small, as shown in simulations 
\citep{XY13}.
The exact scaling relation between $\langle y^2 \rangle$ and $s$ depends on 
$\lambda_\|$ of bouncing CRs (Section \ref{sec: bounc}).
$\lambda_\|$ is related to the parallel diffusion coefficient $D_\|$ by 
$D_\| = 1/3 v \lambda_\|$,
where $v$ is particle velocity. 
$D_\|$ of bouncing CRs
will be derived in the next sections. }


\begin{figure}[htbp]
   \includegraphics[width=7cm]{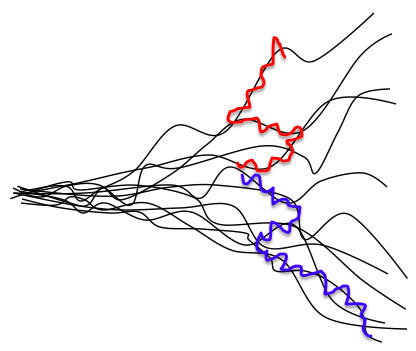}
\caption{
Because of the perpendicular superdiffusion of particles following {\xu turbulent} magnetic field lines, 
particles that undergo bouncing move diffusively along the {\xu local} magnetic field. 
Thin lines represent magnetic field lines. Thick lines represent the trajectories of two particles whose initial separation is small.  }
\label{fig: sket}
\end{figure}

\section{Parallel diffusion of bouncing particles induced by fast modes in compressible MHD turbulence}
\label{sec:bnbfast}

\subsection{Bouncing and non-bouncing particles}

In compressible MHD turbulence, if fast modes carry a significant fraction of the injected turbulent energy, they act as the main agent for scattering particles
\citep{YL02,YL04}. 
More recently, XL20 identified the important role of fast modes in both bouncing and scattering CRs. 
XL20 studied the parallel diffusion associated with the gyroresonant scattering of non-bouncing particles. 
Here we will focus on the parallel mirror diffusion of bouncing particles.

The scattering causes diffusion in pitch angle. 
In the quasilinear approximation
\citep{Jokipii1966},
for the gyroresonant scattering by fast modes, the pitch-angle diffusion coefficient is 
\citep{Volk:1975}
\begin{equation}\label{eq: comgyd}
      D_{\mu\mu,\text{QLT},f} = C_\mu \int d^3k \frac{k_\|^2}{k^2} [J_1^\prime (x)]^2 I_f(k) R(k)  .
\end{equation}
Here 
\begin{equation}\label{eq: cmu}
    C_\mu  = (1 - \mu^2) \frac{\Omega^2}{B_0^2}, ~~ x = \frac{k_\perp v_\perp}{ \Omega}  = \frac{k_\perp }{r_g^{-1} },
\end{equation}
$B_0$ is the mean magnetic field strength, 
$v_\perp$ is the perpendicular component of $v$,
$\Omega$ is the gyrofrequency,
$r_g$ is the particle gyroradius, 
and $k_\|$ and $k_\perp$ are parallel and perpendicular components of wavenumber $k$.
Besides, 
\begin{equation}\label{eq: fsep}
     I_f(k) = C_f k^{-\frac{7}{2}}
\end{equation}
is the energy spectrum of fast modes 
\citep{CL02_PRL},
with
\begin{equation}\label{eq: cf}
    C_f = \frac{1}{16 \pi} \delta B_f^2 L^{-\frac{1}{2}},
\end{equation}
where $\delta B_f$ is the magnetic perturbation of fast modes at $L$, 
and 
\begin{equation}\label{eq: qltreg}
    R = \pi \delta(\omega_k - v_\| k_\| + \Omega )  
\end{equation}
is the resonance function for gyroresonance in the quasilinear approximation, 
where $\omega_k$ is the wave frequency.
We see that when the pitch angle approaches $90^\circ$, the above resonance condition cannot be satisfied with a limited range of $k$ and decreasing magnetic fluctuation amplitude with increasing $k$, leading to the $90^\circ$ problem\footnote{The vanishing scattering close to $90^\circ$ and infinite mean free path in the quasi-linear theory 
is known as the $90^\circ$ problem
\citep{Fisk74}.}.
For our analytical estimate, we can use the approximate expression of $D_{\mu\mu,\text{QLT},f}$
\citep{Xuc16,XLb18}
\begin{equation}\label{eq: appfqltu}
\begin{aligned}
      D_{\mu\mu,\text{QLT},f} 
         &  \approx \frac{ \pi}{56} \frac{ \delta B_f^2}{B_0^2}   \Big(\frac{v}{L\Omega}\Big)^\frac{1}{2}
                \Omega  (1 - \mu^2) \mu^{\frac{1}{2}}  .
\end{aligned}
\end{equation}
We define the rate of change in $\mu$ due to scattering as the scattering rate (XL20),
\begin{equation}\label{eq: scraft}
    \Gamma_{s,f} = \frac{2 D_{\mu\mu,\text{QLT},f}}{\mu^2} \approx
          \frac{ \pi}{28} \frac{ \delta B_f^2}{B_0^2}   \Big(\frac{v}{L\Omega}\Big)^\frac{1}{2}
                \Omega  (1 - \mu^2) \mu^{-\frac{3}{2}}  .
\end{equation}

In the presence of the stochastic magnetic mirrors induced by fast modes, 
the particles that satisfy the bouncing condition (see below) undergo the bouncing motions 
among different magnetic mirrors as discussed in Section \ref{sssec: pardiff}.
For particles with (see Eq. \eqref{mubk})
\begin{equation}\label{eq: moebc}
    \mu   \approx \sqrt{\frac{b_{fk}}{B_0}}, 
\end{equation}
the bouncing is dominated by 
the magnetic mirrors at $k$ with the magnetic perturbation $b_{fk}$
\citep{CesK73},
where 
\begin{equation}\label{eq: fascc}
    b_{fk} = \delta B_f (kL)^{-\frac{1}{4}}
\end{equation}
according to the scaling of isotropic fast modes 
\citep{CL02_PRL}.
The rate of the adiabatic change in $\mu$ due to the spatial variation of magnetic field is
(\citealt{CesK73},XL20)
\begin{equation}\label{eq: gtraft}
\begin{aligned}
   \Gamma_{b,f}  =   \Big |\frac{1}{\mu} \frac{d\mu}{dt} \Big| &=  \frac{v}{2B_0} \frac{1-\mu^2}{\mu}  b_{fk} k  \\
 & = \frac{v}{2L} \frac{\delta B_f^4}{B_0^{4} } \frac{1-\mu^2}{\mu^7}   
\end{aligned}
\end{equation}
at $\mu>\mu_\text{min,f}$, where 
\begin{equation}
     \mu_\text{min,f} = \sqrt{\frac{b_{fk} (r_g)}{B_0}}  = \sqrt{\frac{\delta B_f}{B_0}} \Big( \frac{r_g}{L} \Big)^{\frac{1}{8}},
\end{equation}
and 
$b_{fk} (r_g)$ is the magnetic perturbation at $r_g$.
At $\mu < \mu_\text{min,f}$, the bouncing rate is 
\begin{equation}\label{eq: trfasmu}
\begin{aligned}
     \Gamma_{b,f}  & =   \frac{v}{2B_0} \frac{1-\mu^2}{\mu}  b_{fk}(r_g) r_g^{-1} \\
                                                                  & =  \frac{v}{2 r_g } \frac{ \delta B_f}{B_0} \Big( \frac{r_g}{L} \Big)^{\frac{1}{4}}  \frac{1-\mu^2}{\mu} .
\end{aligned}
\end{equation}

At the balance between scattering and bouncing, i.e., $\Gamma_{s,f} = \Gamma_{b,f}$, one can find the cutoff $\mu$
(Eqs. \eqref{eq: scraft} and \eqref{eq: gtraft}, 
\begin{equation}\label{eq: faucf}
    \mu_c \approx  \bigg[ \frac{14}{\pi} \frac{ \delta B_f^2}{B_0^2} \Big(\frac{v}{ L \Omega}\Big)^\frac{1}{2} \bigg]^\frac{2}{11}, 
\end{equation}
in agreement with the result in XL20. 
We define the particles with 
$\mu<\mu_c$ as ``bouncing particles"
and those with $\mu>\mu_c$ as ``non-bouncing" particles. 
Their motions are dominated by bouncing and scattering, respectively.
As shown in Fig. \ref{fig: mucfas}, $\mu_c$ increases with the CR energy $E_\text{CR}$ until reaching 
its maximum value 
\begin{equation}
   \mu_{c,\text{max}} = \sqrt{\frac{\delta B_f}{B_0+\delta B_f}}. 
\end{equation}
Here as an illustration, we consider CR protons, the magnetic field strength 
$\delta B_f = B_0=3 \mu$G, 
and $L = 30$~pc as the driving scale of interstellar turbulence.

\begin{figure}[htbp]
\centering   
   \includegraphics[width=9cm]{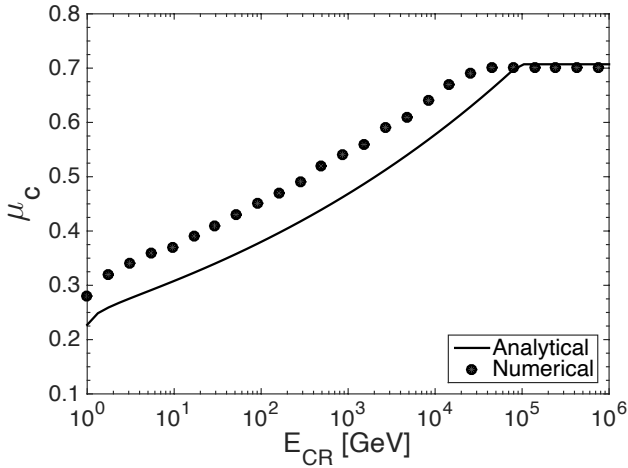}
\caption{$\mu_c$ as a function of $E_\text{CR}$ for fast modes. The analytical approximation is given by Eq. \eqref{eq: faucf}. 
The numerical result is obtained based on the numerical evaluation of Eq. \eqref{eq: comgyd}.}
\label{fig: mucfas}
\end{figure}

\subsection{Parallel diffusion of bouncing particles}
\label{ssec: pdbfast}

When the mirroring condition $\mu < \mu_c$ is satisfied, the parallel diffusion of particles with $\mu$ is regulated by the mirroring effect. 
Based on the relations in Eqs \eqref{eq: moebc} and \eqref{eq: fascc},
bouncing particles diffuse along the magnetic field with a step size 
\begin{equation}\label{eq: mirsz}
     k^{-1} = L \Big(\frac{\delta B_f}{B_0}\Big)^{-4}  \mu^{8}
\end{equation}
for $k^{-1} > r_g$, or equivalently $\mu > \mu_\text{min,f}$.
At $\mu < \mu_\text{min,f}$, the step size of parallel diffusion is $r_g$, corresponding to the Bohm diffusion.  
The maximal step size of bouncing particles corresponds to $\mu_{c}$,
\begin{equation}\label{eq:lambdac}
     k_{c}^{-1}=L \Big(\frac{\delta B_f}{B_0}\Big)^{-4}  \mu_{c}^{8}.
\end{equation}
It decreases with the increase of $\delta B_f / B_0$
(see Eq. \eqref{eq: faucf}) and should not exceed $L$.
The bouncing of the particles with $\mu>\mu_c$ by the magnetic fluctuations on scales larger than $k_c^{-1}$
is less efficient than their gyroresonant scattering by small-scale magnetic fluctuations.

The corresponding parallel diffusion coefficient arising from bouncing is 
\begin{subnumcases}
     { D_{\|,f,b} (\mu)=\label{eq: dufup}}
        v\mu k^{-1} = v L \Big(\frac{\delta B_f}{B_0}\Big)^{-4}  \mu^{9},  \nonumber\\
        ~~~~~~~~~~~~~~~~~~~~~~~~~~~~~\mu_\text{min,f} <\mu <\mu_c ,\\
        v\mu r_g, ~~~~~~~~~~~~~~~~~~~~~~~~~~ \mu < \mu_\text{min,f}.
\end{subnumcases}
The strong dependence on $\mu$ comes from the $\mu$-dependence of the 
size of the dominant magnetic mirror field (Eq. \eqref{eq: mirsz}). 
The bouncing makes the parallel diffusion of CRs with a smaller $\mu$ more inefficient. 
This is opposite to the expectation from the quasi-linear scattering theory.

We first assume an isotropic pitch angle distribution for a simple analytical estimate of the parallel diffusion coefficient. 
By integrating $D_{\|,f,b}(\mu)$ over $\mu$,
we find the parallel diffusion coefficient as a sum of the contributions from 
the multi-scale magnetic mirror field for bouncing particles with different $\mu$,
\begin{equation}\label{eq: simana}
\begin{aligned}
    D_{\|,f,b} &= \int_0^{\mu_c} D_{\|,f,b}(\mu) d\mu \\
         & \approx  \frac{1}{10} v L \Big(\frac{\delta B_f}{B_0}\Big)^{-4} \mu_c^{10}, 
\end{aligned}
\end{equation}
where we consider that the contribution from $D_{\|,f,b}(\mu)$ at $\mu<\mu_\text{min,f}$ is small.

In the situation with an anisotropic pitch angle distribution resulting from the anisotropic scattering 
(see Eq. \eqref{eq: appfqltu}),  
we focus on the pitch-angle diffusion and use the spatially averaged Fokker-Planck equation
\begin{equation}
       \frac{\partial f}{\partial t} = \frac{\partial}{\partial \mu} \bigg[D_{\mu\mu} \frac{\partial f}{\partial \mu} \bigg]
\end{equation}
to derive the particle distribution function $f$ under the effect of scattering. 
If the steady state, i.e., $\partial f/\partial t = 0$,
can be reached, we have 
\begin{equation}
   D_{\mu\mu} \frac{d f}{d \mu} = \text{C}, 
\end{equation}
where $-$C is the constant flux of particles in $\mu$ space. 
The steady-state solution is 
\begin{equation}\label{eq: ssff}
     f(\mu)  = - \text{C}  \int_\mu^{\mu_c} \frac{d\mu^\prime}{D_{\mu\mu} (\mu^\prime)}  + f(\mu_c) .
\end{equation}
To simplify the evaluation of $f(\mu)$, we further adopt the boundary condition
\begin{equation}\label{eq: bocof}
    f (\mu_\text{c}) = 0
\end{equation}
by assuming that 
{\al the diffusion of non-bouncing particles is relatively fast} and 
the non-bouncing particles at $\mu> \mu_c$ instantly escape from the system. 
This assumption implies that the diffusion coefficient of the non-bouncing particles is much larger than that of the bouncing particles. 
{\xu Its validity will be examined later.}
We then normalize the distribution function of the bouncing particles
\begin{equation}
     f^\prime = \frac{f}{ \int_0^{\mu_c} f d\mu}
\end{equation}
to have unit integral, 
\begin{equation}
   \int_0^{\mu_c} f^\prime d\mu = 1.
\end{equation}
The parallel diffusion coefficient of bouncing particles 
under the consideration of anisotropic pitch angle distribution is then
\begin{equation}\label{eq: genfind}
    D_{\|,f,b} = \int_0^{\mu_c} D_{\|,f,b} (\mu) f^\prime d \mu,
\end{equation}
where $D_{\|,f,b}(\mu)$ is given by Eq. \eqref{eq: dufup}. 
It is evident that the above calculation of $D_{\|,f,b}$ for bouncing particles depends on 
$D_{\mu\mu}$ that originates from scattering.

To provide an example for illustrating the parallel diffusion of bouncing particles, 
in Fig. \ref{fig: untrfast}, we present the numerically calculated $D_{\|,f,b} $ as a function of $E_\text{CR}$
and its analytical approximation (Eq. \eqref{eq: simana}). 
The same parameters as used in Fig. \ref{fig: mucfas} are adopted. 
We see that 
the simplification by using an isotropic pitch angle distribution in Eq. \eqref{eq: simana}
does not significantly affect the result. 
This is because 
{\xu for scattering by fast modes},
$D_{\mu\mu,\text{QLT},f}$ and the resulting $f(\mu)$ are not significantly anisotropic.
\footnote{The result derived from isotropic pitch angle distribution is supposed to be larger than the one 
derived from anisotropic distribution. Our analytical approximation underestimates the former due to the 
underestimation of $\mu_c$ (see Fig. \ref{fig: mucfas}).} 
The analytical estimate has the energy dependence as 
$D_{\|,f,b}\propto E_\text{CR}^{10/11}$
for relativistic particles (Eqs. \eqref{eq: faucf} and \eqref{eq: simana}), 
which is close to 
$D_{\|,f,b} \propto E_\text{CR}$. 
The numerical result is a bit shallower and can be fitted by  
$D_{\|,f,b} \propto E_\text{CR}^{0.7}$. 
The discrepancy between the analytical and numerical results at the high-energy end 
comes from the energy dependence of $D_{\|,f,b}(\mu)$ at $\mu<\mu_\text{min,f}$ (Eq. \eqref{eq: dufup}),
which is not taken into account in the analytical approximation.

As a comparison, in Fig. \ref{fig: untrfast} we also add the parallel diffusion coefficient of non-bouncing particles determined by 
the gyroresonant scattering by fast modes derived in XL20. 
The analytical approximation is given by 
\begin{equation}\label{eq: fapasdt}
\begin{aligned}
    D_{\|,f,nb} 
   & =  \frac{v^2}{4 } \int_{\mu_{c}}^{1} d\mu \frac{(1-\mu^2)^2}{D_{\mu\mu,\text{QLT},f}(\mu)}   \\
   &\approx \frac{28}{5\pi} \frac{B_0^2}{ \delta B_f^2} \Big(\frac{v}{L\Omega}\Big)^{-\frac{1}{2}} \frac{v^2}{\Omega}
      \big[4- \sqrt{\mu_{c}} (5-\mu_{c}^2 )\big] ,
\end{aligned}
\end{equation}
where the lower bound in integration is given by $\mu_c$
(see \citealt{CesK73}). 
The resulting $D_{\|,f,nb}$ for non-bouncing particles is shallower than $\propto E_\text{CR}^{0.5}$ 
except for the high-energy end with a constant $\mu_c$.
Its scaling with $E_\text{CR}$ is not an exact power law, but can be approximated by 
$\propto E_\text{CR}^{1/3}$ (see Fig. \ref{fig: untrfast}). 
We see that there is $D_{\|,f,b} < D_{\|,f,nb}$ for all CR energies considered here.  
This justifies the {\xu assumption on the}
boundary condition in Eq. \eqref{eq: bocof}. 
The bouncing particles have a more inefficient diffusion compared with the non-bouncing particles.

We note that in the above illustration, we assume $\delta B_f = B_0$. 
In this case with $\delta B_f < B_0$, 
both bouncing and non-bouncing particles would have larger diffusion coefficients. 
{\al Nevertheless, the diffusion of non-bouncing particles is expected to be faster than that of 
bouncing particles.}

\begin{figure}[htbp]
\centering   
   \includegraphics[width=8.5cm]{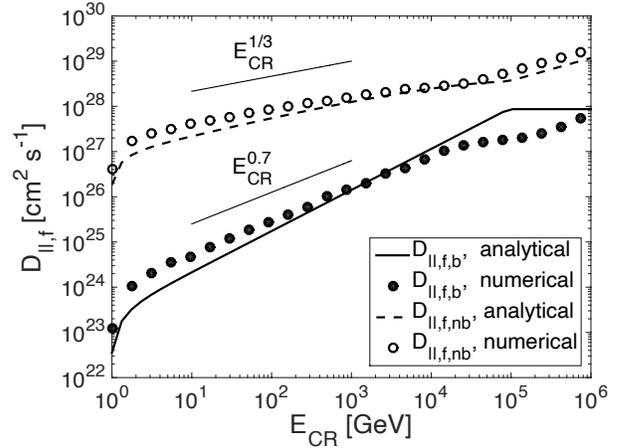}
\caption{Parallel diffusion coefficients $D_{\|,f,b}$ of bouncing CRs 
and $D_{\|,f,nb}$ of non-bouncing CRs as a function of CR energy in compressible MHD turbulence
with fast modes dominating both bouncing and scattering. 
The analytical approximations are given by Eq. \eqref{eq: simana}
and Eq. \eqref{eq: fapasdt}, respectively. }
\label{fig: untrfast}
\end{figure}

\section{Parallel diffusion of bouncing particles in incompressible MHD turbulence} 
\label{sec:incomp}

\subsection{Trans-Alfv\'{e}nic turbulence}
\label{ssec: transa}

{\al CRs propagation in incompressible MHD turbulence was} 
{\xu considered
in many earlier studies using isotropic MHD turbulence model or 2D/slab superposition model 
\citep{Mat90}.}
{\al Later it was found that if turbulent energy is injected at large scales, as this is the case of interstellar turbulence
(e.g., \citealt{ChepH10}),}
the gyroresonant scattering by both Alfv\'{e}n modes and pseudo-Alfv\'{e}n modes is inefficient especially for low-energy CRs
due to the scale-dependent anisotropy of MHD turbulence 
\citep{Chan00,YL02}. 
\citet{YL02}
further argued that 
fast modes in compressible MHD turbulence, 
{\al rather than Alfv\'{e}n modes}, determine the scattering of GeV CRs in our galaxy
{\xu (see \citealt{XLb18} for discussion on importance of slow modes in TTD interaction with CRs)}. 
However, {\al considering the variety of astrophysical conditions,}
{\xu there are situations where fast modes are subject to severe damping 
\citep{YL04,Brunetti_Laz,Xuc16,XLb18}.}
When fast modes are severely damped, or plasma $\beta$ is large
(see Section \ref{ssec: magbet}), or the injected energy of fast modes is small, additional sources for confining CRs are needed. 
{\xu This motives us to} study CR diffusion in {\al idealized} incompressible MHD turbulence or
compressible MHD turbulence with a very small energy fraction in fast modes. {\al In this situation, we focus on the mirroring 
effect} induced by pseudo-Alfv\'{e}n (slow) modes. 
{\al The magnetic mirroring effect has been} found in XL20 to be important for {\xu confining 
CRs in incompressible MHD turbulence. 
Here we will formulate the corresponding parallel diffusion coefficient.}

The pitch-angle diffusion coefficients for gyroresonant interactions 
with Alfv\'{e}n and pseudo-Alfv\'{e}n modes 
in the quasi-linear approximation are 
\citep{Volk:1975},
\begin{equation}\label{eq: oriduav}
     D_{\mu\mu, \text{QLT},A} = C_\mu \int d^3k x^{-2} [J_1(x)]^2 I_A(k) R(k),
\end{equation}
and 
\begin{equation}\label{eq: comgyds}
      D_{\mu\mu,\text{QLT},s} = C_\mu \int d^3k \frac{k_\|^2}{k^2} [J_1^\prime (x)]^2 I_s(k) R(k)  .
\end{equation}
The magnetic energy spectra are  
\citep{CLV_incomp}
\begin{equation}\label{eq: enespa}
     I_A(k) = C_A  k_\perp^{-\frac{10}{3}} \exp{\Bigg(-L^\frac{1}{3}\frac{k_\|}{k_\perp^\frac{2}{3}}\Bigg)},~ 
    C_A = \frac{1}{6 \pi} \delta B_A^2 L^{-\frac{1}{3}}
\end{equation}
for Alfv\'{e}n modes, and 
\begin{equation}\label{eq: slspe}
     I_s(k) = C_s  k_\perp^{-\frac{10}{3}} \exp{\Bigg(-L^\frac{1}{3}\frac{k_\|}{k_\perp^\frac{2}{3}}\Bigg)},~
    C_s = \frac{1}{6 \pi} \delta B_s^2 L^{-\frac{1}{3}}
\end{equation}
for pseudo-Alfv\'{e}n modes,
where $\delta B_A$ and $\delta B_s$ are their magnetic
perturbations at $L$.
The corresponding scattering rate with contributions from both Alfv\'{e}n and pseudo-Alfv\'{e}n modes is 
\begin{equation}
    \Gamma_{s,\text{inc}} = \frac{2(D_{\mu\mu,\text{QLT},A} + D_{\mu\mu,\text{QLT},s})}{\mu^2}.
    \label{eq:Gs}
\end{equation}

Besides scattering, pseudo-Alfv\'{e}n modes also 
generate magnetic mirrors accounting for the bouncing of CRs. 
The particles with 
\begin{equation} \label{eq:mu} 
     \mu \approx \sqrt{\frac{b_{sk}}{B_0}} 
\end{equation}
mainly undergo the bouncing motion at $k_\|$, where the magnetic perturbation of pseudo-Alfv\'{e}n modes is 
\begin{equation}\label{eq: anislber}
    b_{sk} = \delta B_s (k_\perp L)^{-\frac{1}{3}}
               = \delta B_s (k_\| L)^{-\frac{1}{2}}
\end{equation}
based on the anisotropic scaling of MHD turbulence
\citep{CLV_incomp}.
The bouncing rate is
(\citealt{CesK73}, XL20) 
\begin{equation}\label{eq: slbcrlm}
\begin{aligned}
    \Gamma_{b,s} =  \Big| \frac{1}{\mu} \frac{d\mu}{dt} \Big| &=  \frac{v}{2B_0} \frac{1-\mu^2}{\mu}  b_{sk} k_\| \\
  &  =  \frac{v}{2 L } \aleph^2_s \frac{1-\mu^2}{\mu^3}     
\end{aligned}
\end{equation}
at $\mu > \mu_\text{min,s}$, where
\begin{equation}
  \aleph_s  =  \frac{ \delta B_s}{B_0},
\end{equation}
\begin{equation}
   \mu_\text{min,s} = \sqrt{ \frac{b_{sk} (r_g)}{B_0}} = \sqrt{\aleph_s} \Big(\frac{r_g}{ L} \Big)^{\frac{1}{4}},
\end{equation}
and $b_{sk} (r_g)$ is the magnetic perturbation at $k_\| = 1/ r_g$. 
At $\mu < \mu_\text{min,s}$, the bouncing rate becomes 
\begin{equation}\label{eq: trfasmusl}
\begin{aligned}
    \Gamma_{b,s}  & = \frac{v}{2B_0} \frac{1-\mu^2}{\mu}  b_{sk} (r_g) r_g^{-1} \\
                                                                  & = \frac{v}{2 r_g }  \aleph_s \Big(\frac{r_g}{ L} \Big)^{\frac{1}{2}}  \frac{1-\mu^2}{\mu}.
\end{aligned}     
\end{equation}
Here we note that as $\Gamma_{b,s}$ and $\Gamma_{b,f}$ (Eqs. \eqref{eq: gtraft} and \eqref{eq: trfasmu})
have different $\mu$ dependence, 
in the presence of both slow and fast modes in compressible MHD turbulence, 
the relative importance between the mirroring effects induced by slow and fast modes can change with $\mu$
(see Section \ref{ssec: comvarf}).

As found in XL20, 
because of the inefficient scattering by Alfv\'{e}n and pseudo-Alfv\'{e}n modes, 
the effect of magnetic mirrors dominates over the gyroresonant scattering, i.e. $\Gamma_{b,s} > \Gamma_{s,\text{inc}}$,
for $\mu < \mu_c$,
and $\mu_c$ is given by its maximum value
\footnote{$\mu_c$ in incompressible MHD turbulence was approximated to be unity in XL20 for simplicity.}
\begin{equation}\label{eq: maxmuinc}
   \mu_c = \sqrt{\frac{\delta B_s}{B_0 + \delta B_s}},
\end{equation}
which is independent of $E_\text{CR}$.

The non-bouncing particles with $\mu>\mu_c$ are only poorly constrained by the inefficient scattering, 
and their diffusion coefficients are expected to be large (see Section \ref{sec: gencase}). 
For the parallel diffusion of bouncing particles, the step size is 
\begin{equation}
    k_\|^{-1} = L \aleph^{-2}_s \mu^{4}
    \label{eq:kparslow}
\end{equation}
at $\mu > \mu_\text{min,s}$, or equivalently, $k_\|^{-1} > r_g$, 
and the step size is $r_g$ at $\mu < \mu_\text{min,s}$.
Accordingly, the $\mu$-dependent parallel diffusion coefficient is 
\begin{subnumcases}
     { D_{\|,\text{inc},b} (\mu)=\label{eq: duincst}}
         v\mu k_\|^{-1} =  v L \aleph^{-2}_s \mu^5,  \nonumber\\
        ~~~~~~~~~~~~~~~~~~~~~~~~~~~\mu_\text{min,s} <\mu <\mu_c ,\\
         v \mu r_g, ~~~~~~~~~~~~~~~~~~~~~~~~~~~ \mu < \mu_\text{min,s}. \label{eq:duincstsmu}
\end{subnumcases}
If the pitch angle distribution of bouncing particles is isotropic, 
then the parallel diffusion coefficient can be approximated by 
\begin{equation}\label{eq: anainc}
   D_{\|,\text{inc},b} \approx \int_0^{\mu_c} D_{\|,\text{inc},b}(\mu) d\mu =\frac{1}{6} vL  \aleph_s^{-2} \mu_c^6,
\end{equation}
which is energy independent with a constant $\mu_c$. 
Here we neglect the contribution from $D_{\|,\text{inc},b} (\mu)$ at $\mu < \mu_\text{min,s}$.


We next consider the situation with an anisotropic pitch angle distribution. 
We follow the similar analysis in Section \ref{ssec: pdbfast} 
to derive $D_{\|,\text{inc},b}$ numerically by using Eqs. \eqref{eq: ssff}-\eqref{eq: genfind}. 
In Eq. \eqref{eq: ssff}, we adopt (Eqs. \eqref{eq: oriduav} and \eqref{eq: comgyds})
\begin{equation}\label{eq: duuast}
    D_{\mu\mu} (\mu) = D_{\mu\mu,\text{QLT},A} + D_{\mu\mu,\text{QLT},s}
\end{equation}
to include the gyroresonant scattering by both Alfv\'{e}n and pseudo-Alfv\'{e}n modes. 
Eq. \eqref{eq: genfind} becomes 
\begin{equation}\label{eq: intincan}
    D_{\|,\text{inc},b} = \int_0^{\mu_c} D_{\|,\text{inc},b}(\mu) f^\prime d \mu,
\end{equation}
where $D_{\|,\text{inc},b}(\mu)$ is given by Eq. \eqref{eq: duincst}.

In Fig. \ref{fig: trapfast}, we illustrate the numerically calculated $D_{\|,\text{inc},b}$ with an anisotropic pitch angle 
distribution in comparison with the analytical approximation with an isotropic pitch angle distribution
(Eq. \eqref{eq: anainc}).
The same parameters as in Fig. \ref{fig: mucfas} are used, and the turbulence is considered to be 
trans-Alfv\'{e}nic with $\delta B_A = B_0$. Here we also use $\delta B_s = B_0$, i.e., $\aleph_s=1$.
We see that the anisotropic distribution leads to a significantly smaller $D_{\|,\text{inc},b}$ than that derived from an isotropic 
distribution. 
Fig. \ref{fig: appf} displays the anisotropic $f^\prime$ caused by the anisotropic pitch-angle scattering
for different CR energies. 
The highly anisotropic scattering originates from the scale-dependent anisotropy of 
Alfv\'{e}n and pseudo-Alfv\'{e}n modes. 
At a small $\mu$, 
the gyroresonance with 
many uncorrelated eddies with the perpendicular eddy size much smaller than $r_g$
makes the scattering inefficient
\citep{Chan00,YL02}.
Consequently, most particles are concentrated at a small $\mu$.

The more pronounced anisotropy of $f^\prime$ for low-energy CRs is caused by the inner cutoff at $k_\text{max}$
of turbulent energy spectrum. 
XL20 found that for the gyroresonance with 
Alfv\'{e}n and pseudo-Alfv\'{e}n modes, the magnetic fluctuations at 
\begin{equation}
   k_{\perp p} = \Big(\frac{L^\frac{1}{3}k_{\|,\text{res}}}{8}\Big)^\frac{3}{2} = 8^{-\frac{3}{2}}k_{\perp,\text{res}}
\end{equation}
play a dominant role in determining the scattering efficiency.
Here the parallel and perpendicular resonant wavenumbers are 
$k_{\|,\text{res}} \approx \Omega / v_\|$ and $k_{\perp,\text{res}}= L^\frac{1}{2} k_{\|,\text{res}}^\frac{3}{2}$ for 
trans-Alfv\'{e}nic turbulence. 
The former is derived from the resonance condition in Eq. \eqref{eq: qltreg}. 
When $k_{\perp p} > k_\text{max}$ at a small $\mu$ and a low $E_\text{CR}$, 
the scattering becomes ineffective,
leading to a more anisotropic $f^\prime$. 
As a result, the integral in Eq. \eqref{eq: intincan} is dominated by $D_{\|,\text{inc},b}(\mu)$ at a small $\mu$
(Eq. \eqref{eq:duincstsmu}) (see Fig. \ref{fig: appprod}). 
For the parameters adopted here, 
the CR energy corresponding to $k_{\perp p} =k_\text{max} =10^{-8}$ cm$^{-1}$ and the minimum $\mu=0.01$ 
is indicated by the dashed line in Fig. \ref{fig: trapfast}, 
below which the damping of turbulent spectrum has a significant effect on the diffusion of bouncing particles. 
The enhanced anisotropy due to damping gives rise to more suppressed diffusion.

At higher energies, Eq. \eqref{eq: intincan} is dominated by 
$D_{\|,\text{inc},b}(\mu)$ at a larger $\mu$, where both $f^\prime$ and $D_{\|,\text{inc},b}(\mu)$ remain unchanged
(Figs. \ref{fig: appf} and \ref{fig: appmu}), 
and the resulting $D_{\|,\text{inc},b}$ becomes energy independent (Fig. \ref{fig: trapfast}).
For high-energy CRs, Eq. \eqref{eq: intincan} is again dominated by $D_{\|,\text{inc},b}(\mu)$ at a small $\mu$ similar to 
low-energy CRs, 
but due to the increase of $r_g$ with $E_\text{CR}$ (Eq. \eqref{eq:duincstsmu}). 
At both low- and high-energy ends, $D_{\|,\text{inc},b}$ has the energy dependence close to 
$D_{\|,\text{inc},b} \propto E_\text{CR}$ as dictated by Eq. \eqref{eq:duincstsmu}.

In the vicinity of a CR source, 
the initial pitch angle distribution of the injected CRs 
is important for determining the parallel diffusion of bouncing CRs 
until they lose the memory about the initial distribution via scattering. 
Therefore, the CRs closer to the source can have slower diffusion. 
{\xu We note that the pitch angle distribution of low-energy CRs 
can also be affected by CR-driven instabilities 
especially near CR sources. Intensive scattering by Alfv\'{e}n waves generated by streaming CRs can reduce the anisotropy in pitch angle distribution.
This effect should be taken into account for more realistic modeling of the paralell diffusion coefficient of low-energy CRs
(see, e.g., \citealt{Bla12}).}


We see that even though the pitch-angle scattering is inefficient in incompressible 
trans-Alfv\'{e}n MHD turbulence, 
the mirroring effect due to the presence of pseudo-Alfv\'{e}n modes 
significantly suppresses the parallel diffusion of CRs.

\begin{figure}[htbp]
\centering   
   \includegraphics[width=9cm]{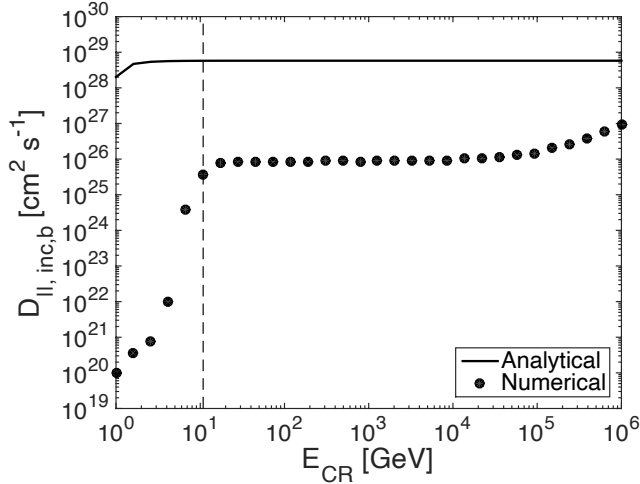}
\caption{Parallel diffusion coefficient $D_{\|,\text{inc},b}$ of bouncing CRs as a function of CR energy in incompressible MHD turbulence. The numerical and analytical (Eq. \eqref{eq: anainc}) results are obtained using 
anisotropic and isotropic pitch angle distribution, respectively. 
The dashed line indicates the energy, below which the diffusion is affected by damping of turbulence.  }
\label{fig: trapfast}
\end{figure}

\begin{figure*}[htbp]
\centering   
\subfigure[]{
   \includegraphics[width=8.5cm]{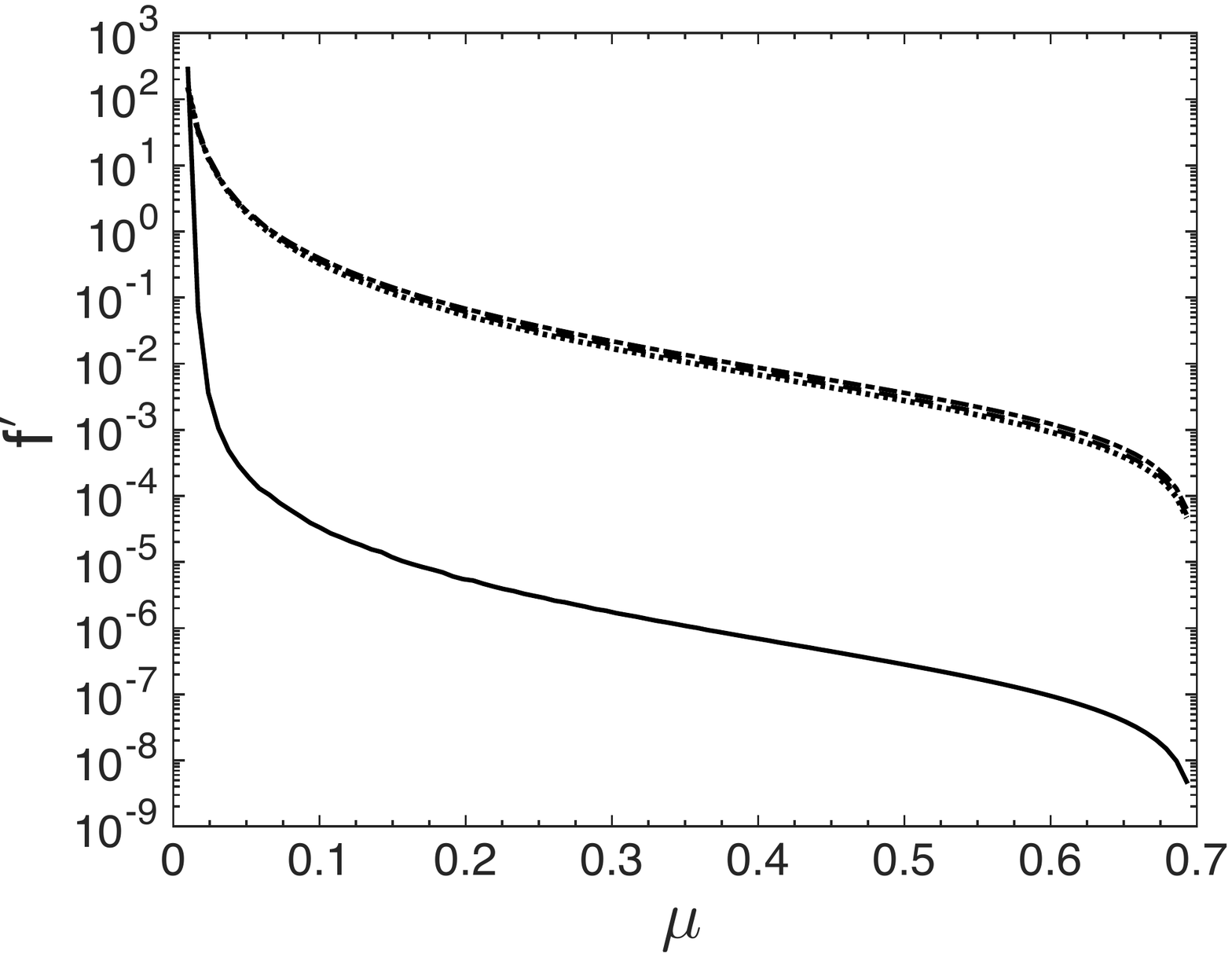}\label{fig: appf}}
\subfigure[]{
   \includegraphics[width=8.5cm]{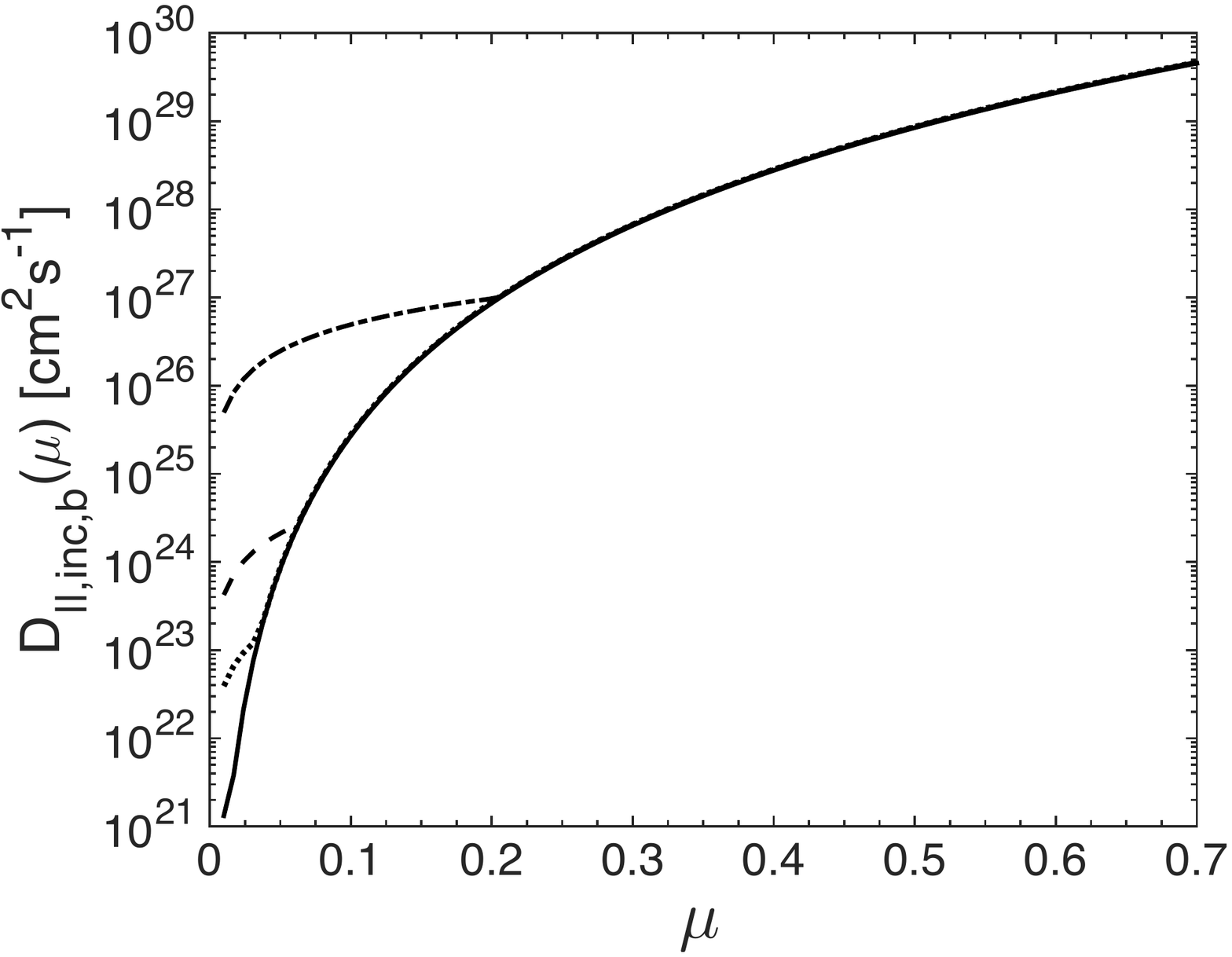}\label{fig: appmu}}
\subfigure[]{
   \includegraphics[width=8.5cm]{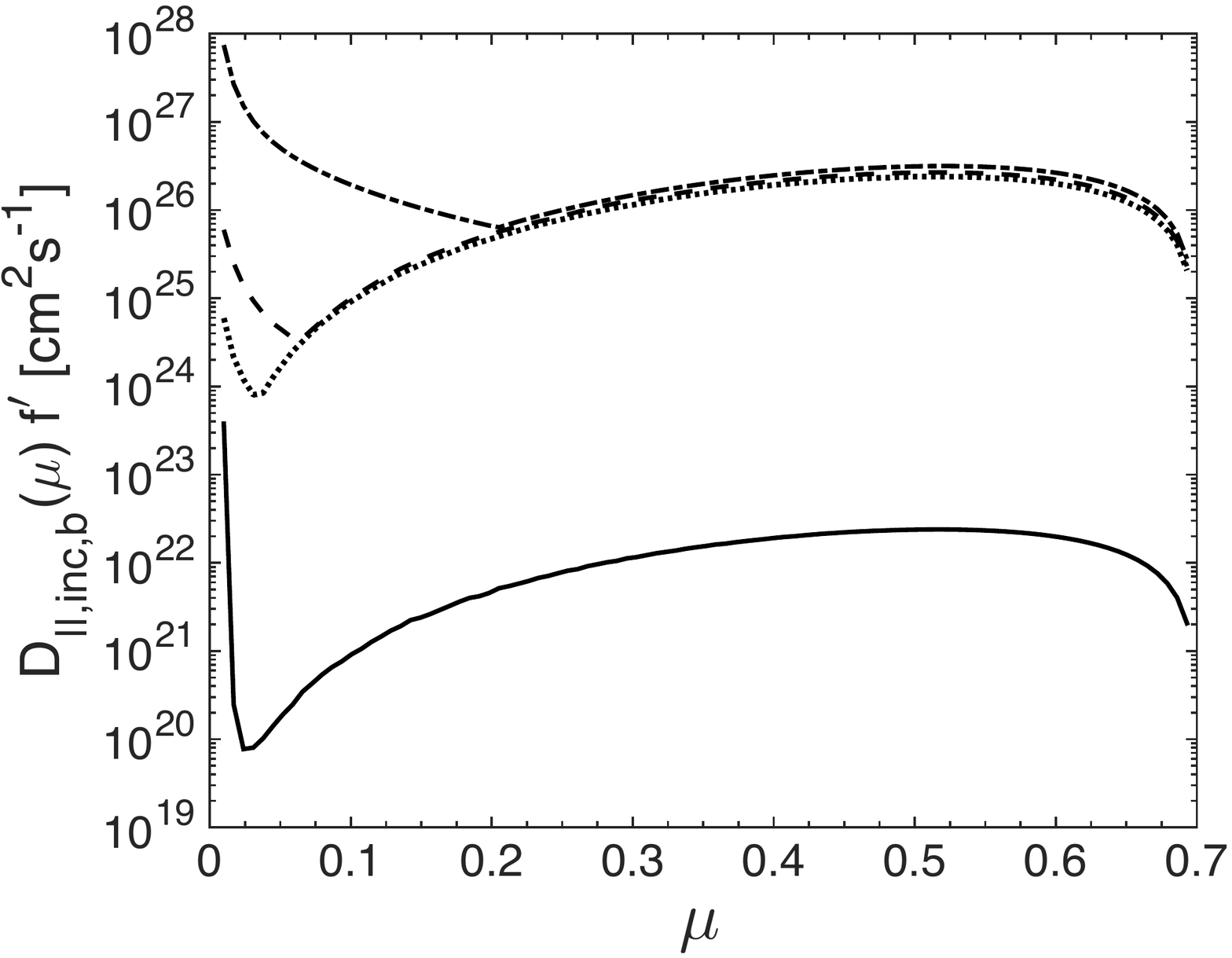}\label{fig: appprod}}
\caption{ (a) Normalized particle distribution function $f^\prime$, (b) $\mu$-dependent parallel diffusion coefficient $D_{\|,\text{inc},b}(\mu)$, 
and (c) $D_{\|,\text{inc},b}(\mu) f^\prime$ vs. $\mu$ for 
$4$ GeV (solid line), $10^2$ GeV (dotted line),
$10^3$ GeV (dashed line), and $10^5$ GeV (dash-dotted line) bouncing CRs 
in incompressible MHD turbulence.}
\label{fig: slottd}
\end{figure*}

\subsection{Super- and sub-Alfv\'{e}nic turbulence}

In the above analysis, we discussed the case of trans-Alfv\'{e}nic turbulence
with $M_A =1$, or equivalently, $\delta B_A = B_0$. 
Our analysis can be generalized for both super- and sub-Alfv\'{e}nic turbulence.

For super-Alfv\'{e}nic turbulence ($M_A>1$), 
the perpendicular superdiffusion
\citep{LV99,  YL08, Lazarian06,LY14}
takes place on scales less than $l_A$.  
\footnote{On scales larger than $l_A$, 
CRs following magnetic field lines undergo isotropic diffusion with the step size equal to $l_A$ 
(see \citealt{Brunetti_Laz}).}
For the analysis in Section \ref{ssec: transa} for anisotropic MHD turbulence to be applicable, 
we use $l_A$ (Eq. \eqref{eq:A12}) as the effective injection scale of turbulence 
with the injected turbulent velocity equal to $V_A$. 
Therefore, after replacing $L$ by $l_A$, we can still use the expressions in Section \ref{ssec: transa}
for describing the parallel diffusion of bouncing particles on scales smaller than $l_A$ in 
super-Alfv\'{e}nic turbulence. 
We note that $\delta B_s$ in this case should be measured at $l_A$ instead of $L$.

In sub-Alfv\'{e}nic turbulence with $M_A <1$,
as we discuss in Appendix A, the so-called weak turbulence 
\citep{LV99,Gal00}
exists on scales from $L$ down to $l_\text{tran}$. 
Strong MHD turbulence is only developed on scales below $l_\text{tran}$. 
In the strong turbulence regime in sub-Alfv\'{e}nic turbulence, 
the turbulence scaling is different from that in trans-Alfv\'{e}nic turbulence, 
and the turbulent eddies are more elongated, 
with the elongation depending on $M_A$
\citep{LV99}. 
We notice in Appendix A, if we introduce an effective injection scale $L_\text{eff}$ for sub-Alfv\'{e}nic turbulence that is given by Eq. \eqref{eq:A11}, then we can describe the sub-Alfv\'{e}nic turbulence at scales less than $l_\text{tran}$ as arising from the fictitious driving with the injection scale $L_\text{eff}$ and injection velocity $V_A$.
$L_\text{eff}$ is larger than the actual injection scale $L$ by a factor of $M_A^{-4}$. 
Unlike $l_A$ in super-Alfv\'{e}nic turbulence, where the transition from isotropic to 
anisotropic turbulence occurs, 
$L_\text{eff}$ does not have a well-defined physical meaning. 
Its introduction, nevertheless, allows us to generalize the results for trans-Alfv\'{e}nic turbulence 
to the strong turbulence regime of sub-Alfv\'{e}nic turbulence by replacing $L$ used in Section \ref{ssec: transa}
by $L_\text{eff}$. 
$\delta B_s$ in this case should be measured at $L_\text{eff}$, and 
it is related to the magnetic perturbation $\delta B_{s,L}$ of pseudo-Alfv\'{e}n modes at $L$ by 
\begin{equation}
    \delta B_s=\delta B_{s,L} M_A^{-2}.
\end{equation}
In Table \ref{tab:gen}, we summarize the 
parameters introduced for generalizing the results on diffusion of bouncing particles in trans-Alfv\'{e}nic turbulence
to super- and sub-Alfv\'{e}nic turbulence.

\begin{table*}[!htbp]
\renewcommand\arraystretch{1.5}
\centering
\begin{threeparttable}
\caption[]{}\label{tab:gen} 
  \begin{tabular}{c|c|c|c}
     \toprule
                                        &   $L_\text{eff}$                  &   $\delta B_s$ at $L_\text{eff}$ &          Range of length scales \\
                                                      \hline
    Super-Alfv\'{e}nic turbulence  ($M_A>1$)   &    $l_A = L M_A^{-3}$        &$\delta B_s$~at $l_A$ &               $<l_A$ \\
    Sub-Alfv\'{e}nic turbulence ($M_A<1$)      &    $L M_A^{-4}$                   &  $\delta B_{s,L} M_A^{-2}$ &             $< L M_A^2$ \\
     \bottomrule
    \end{tabular}
 \end{threeparttable}
\end{table*}

\section{Averaged parallel diffusion coefficients 
{\xu on scales $\gg \lambda_\|$}}
\label{sec: gencase}


The analysis on the parallel diffusion of bouncing particles in Sections \ref{sec:bnbfast} and \ref{sec:incomp}
can be applied to the bouncing CRs near a CR source. 
On scales much larger than $\lambda_\|$ of both bouncing and non-bouncing particles, 
we need to consider their exchange  
and the isotropization of pitch angle distribution due to scattering. 
In this case we define an averaged parallel diffusion coefficient for all particles as 
\begin{equation}\label{eq: avedpa}
    D_{\|,\text{tot}}\approx 
          \alpha D_{\|,b} + (1-\alpha) D_{\|,nb} ,
\end{equation}
where 
\begin{equation}
    \alpha = \frac{\tau_b}{\tau_b+\tau_{nb}}, ~~ 1-\alpha = \frac{\tau_{nb}}{\tau_b+\tau_{nb}} , 
\end{equation}
$\tau_{b}$ and $\tau_{nb}$ represent the times for particles to stay in the 
bouncing state and non-bouncing state, respectively. 
As a simple estimate, we have 
\begin{equation}
    \tau_b\approx \frac{\mu_c^2}{D_{\mu\mu}},~~\tau_{nb}\approx \frac{1-\mu_c^2}{D_{\mu\mu}}
\end{equation}
to account for the diffusion in pitch angle by scattering 
and the resulting transition from bouncing (non-bouncing) to non-bouncing (bouncing) particles. 
For the weakly anisotropic pitch angle distribution under consideration, 
we can use $D_{\mu\mu}$ at $\mu_c$ in the above expression as an approximation. 
Therefore, Eq. \eqref{eq: avedpa} can be rewritten as 
\begin{equation}\label{eq: dtave}
   D_{\|,\text{tot}}\approx \mu_c^2 D_{\|,b} + (1-\mu_c^2) D_{\|,nb} .
\end{equation}

Over a timescale $T$ much longer than $\tau_{b}$ and $\tau_{nb}$, 
the mean squared displacement of CRs in the bouncing state is 
\begin{equation}
    \Delta_b^2 = D_{\|,b}\alpha T,
\end{equation}
and the mean squared displacement of CRs in the non-bouncing state is  
\begin{equation}
    \Delta_{nb}^2 = D_{\|,nb}(1-\alpha) T. 
\end{equation}
The total mean squared 
displacement $\Delta_\text{tot}^2=\Delta_b^2+\Delta_{nb}^2$ during $T$ corresponds to 
a total diffusion coefficient  
\begin{equation}
D_{\|, \text{tot}}= \frac{\Delta_\text{tot}^2}{T}\approx \alpha D_{\|,b}+
(1-\alpha) D_{\|,nb},
\label{Dtot}
\end{equation}
which recovers Eq. \eqref{eq: avedpa}. 
In the case with $D_{\|,b} \ll D_{\|,nb}$,
$D_{\|, \text{tot}}$ can be dominated by the diffusion of non-bouncing particles, that is,
 \begin{equation}
    D_{\|, \text{tot}}\approx (1-\alpha) D_{\|,nb}.  
 \end{equation}
 It is different from the parallel diffusion coefficient in XL20 by a factor of $1-\alpha$. 

\subsection{Compressible MHD turbulence with fast modes dominating magnetic fluctuations}

In Section \ref{sec:bnbfast}, we 
derived the parallel diffusion coefficients of bouncing and non-bouncing particles separately (see Fig. \ref{fig: untrfast}). 
Based on these calculations and 
by using Eq. \eqref{eq: dtave}, here we present $D_{\|,f,\text{tot}}$ for all particles interacting with fast modes 
in compressible MHD turbulence,
as shown in Fig. \ref{fig: totfast}. 
By comparing Fig. \ref{fig: totfast} with Fig. \ref{fig: untrfast}, we see that 
as $D_{\|,f,nb}$ is considerably larger than $D_{\|,f,b}$ for the range of $E_\text{CR}$ under consideration, 
the resulting $D_{\|,f,\text{tot}}$ is determined by $D_{\|,f,nb}$ alone 
and slightly smaller than $D_{\|,f,nb}$ with the increase of $\mu_c$ toward high energies. 
We note that unlike $D_{\|,f}$ usually defined for pitch-angle scattering that corresponds to the change of pitch angle by 
$90^\circ$, 
$D_{\|,f,nb}$ corresponds to a smaller change of pitch angle 
and is smaller than $D_{\|,f}$ with $\mu_c = 0$ (see Eq. \eqref{eq: fapasdt}).

\begin{figure}[htbp]
\centering  
   \includegraphics[width=9cm]{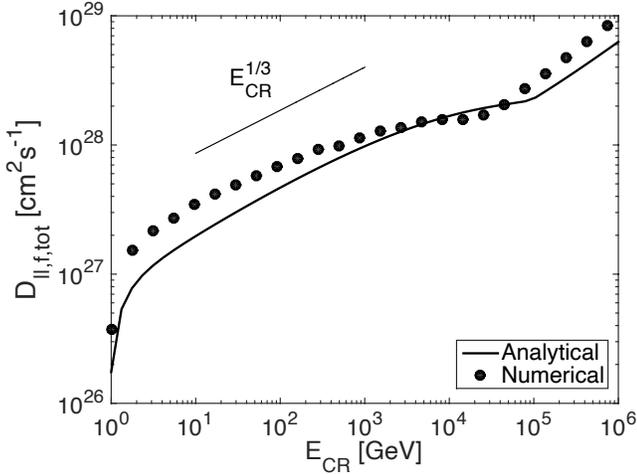}
\caption{
Total parallel diffusion coefficient $D_{\|,f,\text{tot}}$ as a function of $E_\text{CR}$ in compressible MHD turbulence
with magnetic fluctuations dominated by fast modes. }
\label{fig: totfast}
\end{figure}

\subsection{Compressible MHD turbulence with a varying fraction of fast modes}
\label{ssec: comvarf}

In realistic astrophysical media, the energy fraction of fast modes depends on the driving condition of turbulence 
and plasma $\beta$ (see Section \ref{ssec: magbet}). 

In the presence of both fast and slow modes, the ratio between their bouncing rates (Eqs \eqref{eq: gtraft} and \eqref{eq: slbcrlm}) is 
\begin{equation}\label{eq: fstrrat}
    \frac{\Gamma_{b,f}}{\Gamma_{b,s}} = \frac{\delta B_f^4}{B_0^2 \delta B_s^2} \mu^{-4}, 
\end{equation}
when $\mu > \mu_\text{min,f}$ and $\mu > \mu_\text{min,s}$. 
We use the larger one, i.e., 
max$[\Gamma_{b,f},\Gamma_{b,s}]$, 
as the bouncing rate. 
In the case of $\delta B_f \sim \delta B_s \sim B_0$, there is always $\Gamma_{b,f} > \Gamma_{b,s}$
and fast modes dominate bouncing.

{\xu As illustrative examples,}
in Fig. \ref{fig: varyfast}, we consider the parallel diffusion coefficients of bouncing and non-bouncing particles in
trans-Alfv\'{e}nic turbulence with a varying fraction of fast modes, and 
we adopt an isotropic pitch angle distribution. 
$\aleph_s$ is fixed at $0.5$ 
to address the role of fast modes in affecting the diffusion coefficients.


(a) $\delta B_f =0$.
In the limit case of incompressible MHD turbulence with $\delta B_f = 0$
(Fig. \ref{fig: f0s05}), scattering is inefficient and 
$\mu_c$ is given by Eq. \eqref{eq: maxmuinc} (see Fig. \ref{fig: mucnfns}). 
$D_{\|,b}$ is consistent with Eq. \eqref{eq: anainc}.
$D_{\|,nb}$ is 
\begin{equation}
    D_{\|,nb} = \frac{v^2}{4} \int_{\mu_c}^1 d\mu \frac{(1-\mu^2)^2}{D_{\mu\mu}(\mu)} ,
\end{equation}
where $D_{\mu\mu}(\mu)$ is given by Eq. \eqref{eq: duuast}. 
As both $D_{\mu\mu,\text{QLT},A}$ and $D_{\mu\mu,\text{QLT},s}$ have energy dependance as 
$\propto E_\text{CR}^{3/2}$ (XL20), 
the resulting $D_{\|,nb}$ decreases with increasing energy following  
$\propto E_\text{CR}^{-3/2}$.
Because the turbulence anisotropy is weaker toward larger length scales, 
higher-energy CRs are more efficiently scattered and have smaller $D_{\|,nb}$.

(b) $\delta B_f = 0.01 B_0$.
In compressible MHD turbulence, 
we consider the scattering rate with contributions from all three modes, 
\begin{equation}
    \Gamma_{s,\text{tot}} = \frac{2(D_{\mu\mu,\text{QLT},A} + D_{\mu\mu,\text{QLT},s} + D_{\mu\mu,\text{QLT},f})}{\mu^2}.
\end{equation}
When there is a tiny fraction of fast modes with $\delta B_f \ll B_0$ (see Fig. \ref{fig: f001s05}), 
the scattering of low-energy CRs is dominated by fast modes 
despite their small energy fraction. 
With a constant $\mu_c$ (Fig. \ref{fig: mucnfns}), 
$D_{\|,nb}$ determined by $D_{\|,f,nb}$ increases with energy as 
$\propto E_\text{CR}^{0.5}$ (see Eq. \eqref{eq: fapasdt}). 
The scattering of higher-energy CRs is taken over by Alfv\'{e}n and slow modes. 
For bouncing CRs, 
$D_{\|,b}$ is determined by slow modes and 
is the same as that in the incompressible turbulence.

(c) $\delta B_f =0.1 B_0$.
When we further increase the fraction of fast modes (Fig. \ref{fig: f01s05}),
scattering is dominated by fast modes for most energies, 
resulting in the energy dependence of $\mu_c$ until it reaches its maximum value (Fig. \ref{fig: mucnfns}). 
With increasing $\mu_c$, 
$D_{\|,nb}$ has a weaker dependence on $E_\text{CR}$ than $\propto E_\text{CR}^{0.5}$ (Eq. \eqref{eq: fapasdt})
for low-energy CRs. 
For bouncing of CRs, 
although $\Gamma_{b,f}$ is larger than $\Gamma_{b,s}$ at a small $\mu$ for 
low-energy CRs (Eq. \eqref{eq: fstrrat}), 
$D_{\|,b}$ is still determined by slow modes for most energies and 
Eq. \eqref{eq: anainc} applies.

(d) $\delta B_f = 0.5 B_0$.
With a large fraction of fast modes, 
fast modes dominate both scattering and bouncing. 
$D_{\|,nb}$ (see Eq. \eqref{eq: fapasdt}) approximately follows $\propto E_\text{CR}^{1/3}$ (see Fig. \ref{fig: f05s05}). 
$D_{\|,b}$ in this case is given by Eq. \eqref{eq: simana} and is close to $\propto E_\text{CR}$.

The thick solid lines in all cases represent the averaged $D_{\|,\text{tot}}$ (Eq. \eqref{eq: dtave}). 
It is basically determined by $D_{\|,nb}$ as $D_{\|,nb}$ is considerably larger than $D_{\|,b}$.

Based on the above results, our main findings are: 

(i) The difference between $D_{\|,nb}$ dominated by incompressible modes and fast modes 
comes from not only the different slopes of turbulent energy spectra, 
but also the anisotropy of turbulence. 
The scale-dependent turbulence anisotropy in the former case results in a decreasing 
$D_{\|,nb}$ with increasing $E_\text{CR}$.

(ii) Even with a small fraction of fast modes, scattering becomes much more 
efficient than that in incompressible turbulence.

(iii) With an energy-dependent $\mu_c$, $D_{\|,nb}$ dominated by fast modes has a 
weaker dependence on $E_\text{CR}$
than the case with a constant $\mu_c$.

(iv) With a non-zero $\mu_c$ and a factor $1-\alpha$, 
$D_{\|,nb}$ is in general smaller than the parallel diffusion coefficient for gyroresonant scattering with $\mu_c = 0$ 
and $\alpha = 0$.

(v) The maximum $D_{\|,b}$ depends on the maximum $\mu_c$, which is determined by the 
amplitude of magnetic fluctuations ($\delta B_s$ or $\delta B_f$) of the modes that dominate bouncing. 

{\xu (vi) On scales larger than $\lambda_\|$ of both bouncing and non-bouncing particles, the averaged parallel diffusion coefficient is determined by that of non-bouncing particles.}

(vii) The {\xu energy scaling of} $D_{\|,\text{tot}}$ is non-universal, 
{\xu depending on the energy fractions of different turbulence modes.
In diverse astrophysical environments,
the energy fractions of turbulence modes
vary with the turbulence parameters, e.g., $M_s$, $M_A$
\citep{CL03}.
For realistic modeling of CR diffusion, 
prior knowledge of turbulence parameters is necessary, which can be measured with the Gradient Techniques 
(e.g., \citealt{Lazgt18}). 
This falls beyond the scope of this work 
and is left to future studies. }

\begin{figure*}[htbp]
\centering   
\subfigure[$\delta B_f /B_0 = 0$, $\aleph_s = 0.5$]{
   \includegraphics[width=8.5cm]{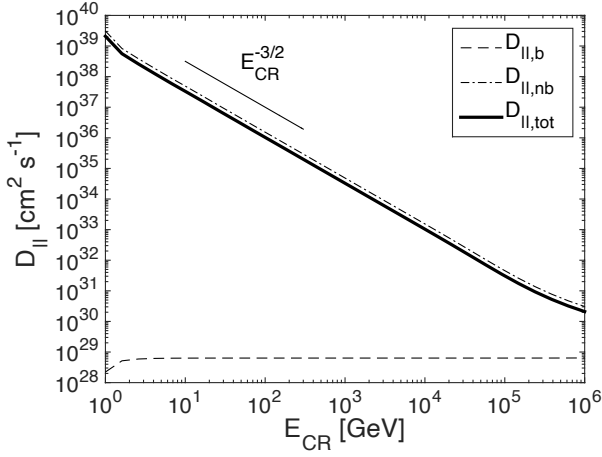}\label{fig: f0s05}}
\subfigure[$\delta B_f /B_0 = 0.01$, $\aleph_s = 0.5$]{
   \includegraphics[width=8.5cm]{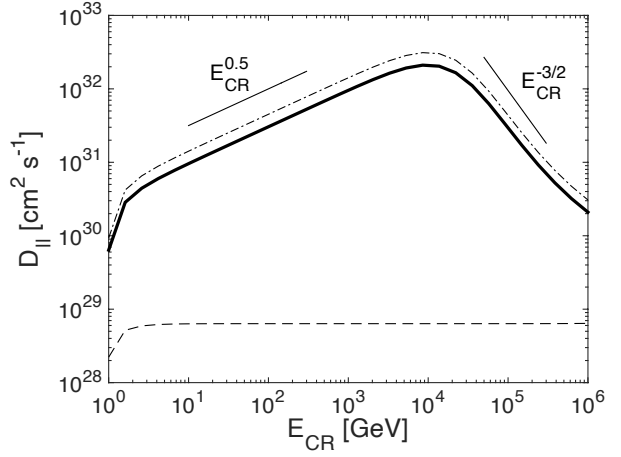}\label{fig: f001s05}}
\subfigure[$\delta B_f /B_0 = 0.1$, $\aleph_s = 0.5$]{
   \includegraphics[width=8.5cm]{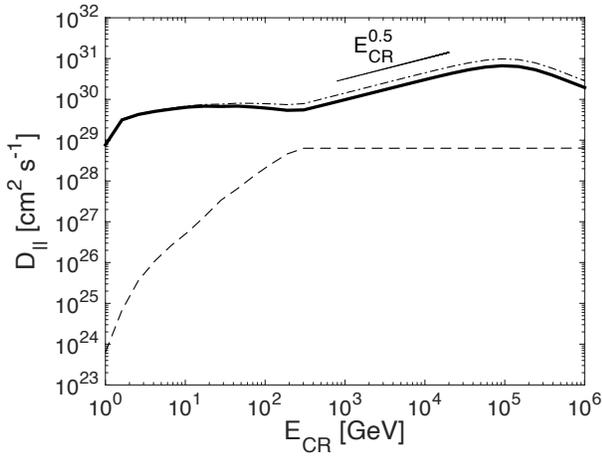}\label{fig: f01s05}}
\subfigure[$\delta B_f /B_0 = 0.5$, $\aleph_s = 0.5$]{
   \includegraphics[width=8.5cm]{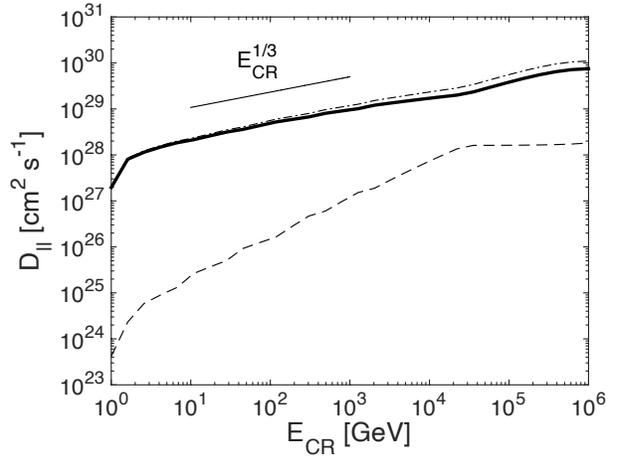}\label{fig: f05s05}}
\caption{Parallel diffusion coefficients of bouncing particles $D_{\|,b}$, non-bouncing particles $D_{\|,nb}$, 
and the averaged parallel diffusion coefficient $D_{\|,\text{tot}}$ in MHD turbulence with 
different fractions of fast modes.
The pitch angle distribution is assumed to be isotropic. }
\label{fig: varyfast}
\end{figure*}

\begin{figure}[htbp]
\centering   
   \includegraphics[width=8cm]
   {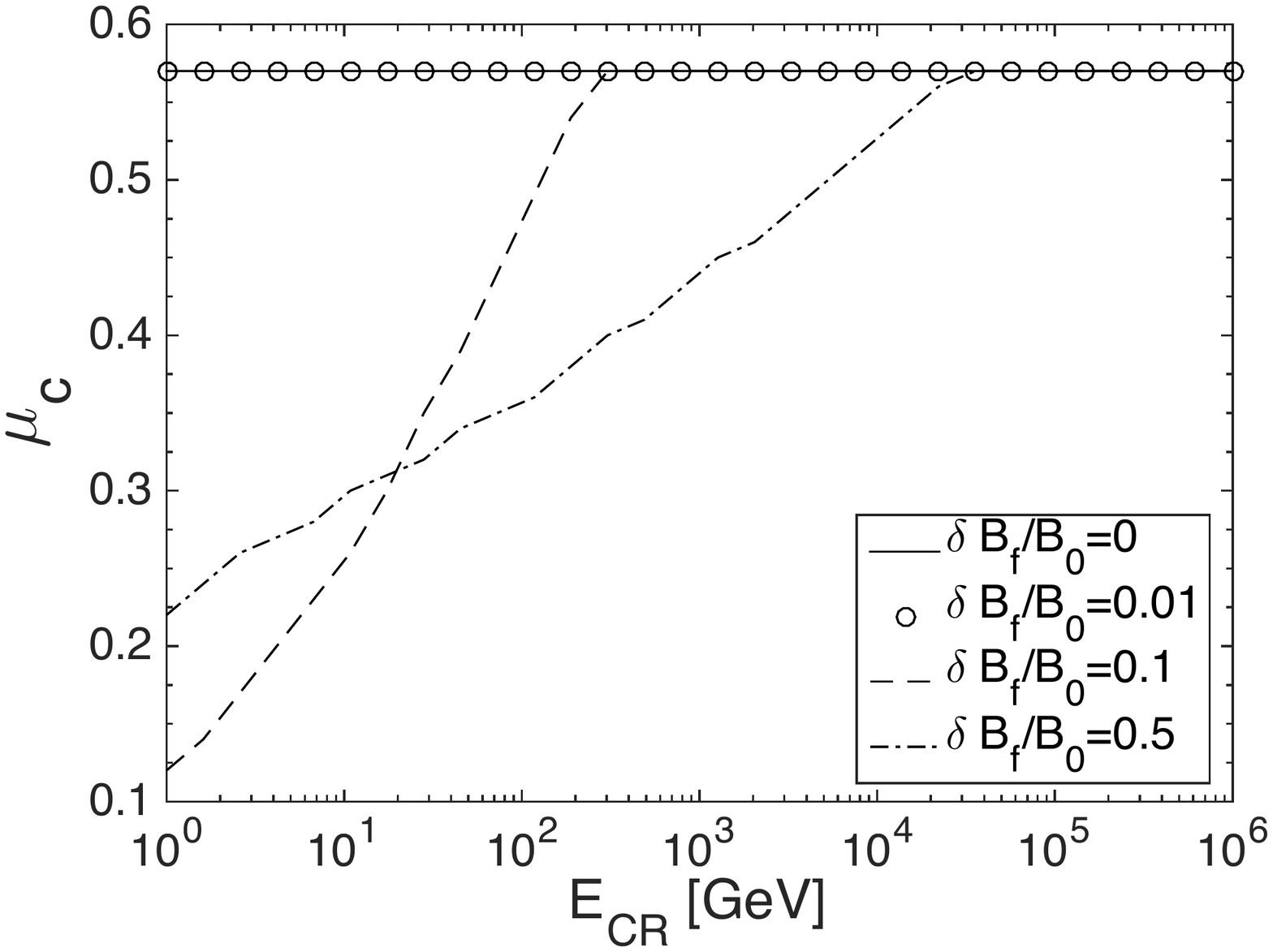}
\caption{ $\mu_c$ vs. $E_\text{CR}$ for different fractions of fast modes. }
\label{fig: mucnfns}
\end{figure}

\section{Discussion}

\subsection{\xu Toward a more comprehensive description of CR propagation}


{\xu Important new findings on CR 
propagation have been reported} 
over the past two decades. 
First of all, the fast modes of MHD turbulence were identified as the major scattering agent for diffusion of CRs with $r_g \ll L$ in the interstellar medium
\citep{YL02,YL03}.
On the contrary, the Alfv\'{e}n and slow modes, which were traditionally considered for CR diffusion, 
were found to be very inefficient in scattering
\citep{Chan00,YL02}.
Therefore, it was believed that, 
when the CR streaming instability is inefficient, e.g. due to Alfvenic turbulence suppression 
(see \citealt{YL02,FG04,La16}),
and fast modes are absent, 
CRs propagate ballistically along magnetic field lines 
without scattering.
\footnote{We note that in super-Alfv\'{e}nic turbulence,
CRs that 
follow the tangled magnetic fields without scattering have an effective mean free path given by $l_A$
\citep{Brunetti_Laz}.}

In this work, we introduce a new diffusion mechanism for CRs that encounter magnetic compressions induced by slow (pseudo-Alfv\'{e}n) and fast modes. 
This {\xu parallel mirror diffusion} of CRs can take place even in incompressible MHD turbulence 
and cause slower diffusion than that induced by scattering in both incompressible and compressible MHD turbulence. 
{\al In the idealized situation with incompressible turbulence and the streaming instability suppressed by Alfv\'{e}nic turbulence, 
bouncing CRs can stay in the system for an extended period of time until they become non-bouncing CRs due to 
scattering or acceleration processes.
This finding has important implications for CR diffusion in high-$\beta$ intracluster media.}
In addition, 
the slow diffusion of bouncing CRs near CR sources may provide
a possible explanation for the small diffusion coefficient of high-energy electrons and positrons {\xu around} pulsar wind nebulae suggested by recent observations 
(e.g., \citealt{Abey17,Hua18}), 
{\xu as well as the steep high-energy CR spectra indicated by gamma-ray observations of middle-aged supernova remnants 
(\citealt{Xu21};
{see also, e.g., \citealt{Nav16,Dang16,Dang18,Nav19},
for suppressed diffusion of self-confined CRs around Galactic sources}}).

In addition, the mirroring of CRs also affects the parallel diffusion associated with scattering. 
In the presence of mirroring, the scattering even in the 
quasilinear approximation does not face the $90^\circ$ problem 
\citep{Noe68,Kulsrud_Pearce,Fel01}, 
and also the resulting mean free path corresponds to 
the pitch angle change over a smaller range $[\mu_c,1]$, 
as pointed out by 
\citet{CesK73}.
In XL20, we derived the parallel diffusion coefficient for scattering by fast modes in the presence of bouncing. 
With an energy-dependent $\mu_c$, 
the diffusion coefficient has a energy dependence
close to $\propto E^{1/3}$ instead of $\propto E^{0.5}$.
In earlier studies (e.g., \citealt{Ptus06}), 
the energy dependence 
$\propto E^{1/3}$ of diffusion coefficient is only expected for the 
Kolmogorov spectrum of turbulence, 
but the Alfv\'{e}n and slow modes with the Kolmogorov spectrum are incapable of scattering.
Our finding provides a {\al possible} solution to this theoretical problem
{\xu and a possible explanation for the energy scaling of diffusion coefficient of high-energy CRs indicated by observations 
(e.g., \citealt{Bla12,Forn21}).}

\subsection{\xu Mirroring and acceleration}

The mirroring of CRs also induces a new acceleration mechanism. 
The mirroring does not change a particle's pitch angle 
if the magnetic mirrors are static. 
When encountering moving magnetic mirrors in MHD turbulence, CRs undergo the second order Fermi acceleration with the stochastic increase of parallel momentum $p_\|$ and $\mu$. 
{\al Together with other acceleration mechanisms, e.g., TTD that also causes the stochastic increase of $\mu$
(Section \ref{ssec:ttd}),}
the mirroring can enhance the diffusion in $\mu$ and 
the transition of particles from bouncing to non-bouncing state. 
This new acceleration mechanism and its implications will be studied in our future work.

The diffusive propagation of CRs plays an important role in the first order Fermi acceleration processes at shocks 
(e.g.,\citealt{Schl17})
and in regions of turbulent magnetic reconnection
\citep{DeG05,Laz05,Kow12}.
Both the perpendicular superdiffusion and the parallel diffusion resulting from mirroring and scattering 
should be taken into account when studying the acceleration of CRs. 

\subsection{\xu Mirroring and instabilities}

{\al Mirroring and resonant scattering by MHD turbulence are not the only processes that affect the diffusion of CRs. The fluid of {\xu low-energy} CRs is subject to instabilities. The streaming instability is the most studied one \citep{Kulsrudbook}. This instability arises when CRs move preferentially in the same direction. The Alfv\'{e}nic fluctuations that are induced are different from MHD turbulence. These are resonant waves that effectively scatter CRs and prevent free escape of particles. }
{\xu Streaming instability can be important for explaining the energy scaling of diffusion coefficient of 
low-energy CRs indicated by AMS-02 measurements
(e.g., \citealt{Forn21}).}

{ \al Apart from the streaming instability, CRs can face other types of instabilities, the importance of which is less explored. For instance, the compression of CRs by turbulence induces anisotropy in the distribution of CRs in momentum space. This can result in the gyroresonance, mirror, and firehose collisionless instabilities of CRs \citep{LB06}. 
{\xu In addition, the faster escape of non-bouncing particles from the vicinity of CRs sounce changes the momentum distribution toward having more CRs with $\mu<\mu_c$.
This difference in the diffusion coefficients} of bouncing and non-bouncing CRs is another source of momentum space anisotropy, which can result in an instability
and induce resonant Alfv\'{e}nic waves. 
The list above does not exhaust the list of the CR instabilities that can be present in the vicinity of CR sources. 
For instance the current instability 
\citep{Bell2004}
was introduced in the context of shock acceleration. 
It is also expected to be present beyond the shock context. 

The instabilities, along with the mirroring, can significantly 
{\xu suppress the diffusion of CRs} 
in the vicinity of their sources. 
The environmental effects, 
e.g. ambient turbulence, gas ionization, have different influence on mirroring and the scattering caused by instabilities. }
We note that 
the CR streaming instability can be suppressed in galactic disks through both ion-neutral collisional  damping 
\citep{Kulsrud_Pearce,Xuc16,Krum20}
and the interaction with Alfv\'{e}nic turbulence 
\citep{YL02,FG04,La16}. 
{\xu More severe damping of streaming instability is expected at a larger $M_A$
\citep{La16}, 
while the {\al magnetic compressions} that cause mirroring increase with $M_A$
(Hu, Lazarian, \& Xu, in prep).
Therefore, the relative importance between streaming instability and mirroring effect on diffusion of low-energy CRs 
depends on $M_A$, which varies in the multi-phase interstellar medium 
\citep{Lazgt18}.
The comparison between scattering by instabilities and mirroring, as well as their interplay,
requires further studies.}


\subsection{\xu Mirroring and TTD}
\label{ssec:ttd}

{\al The TTD effect is an important component of CR dynamics 
\citep{Schlickeiser02, YL02,XLb18}. 
The TTD interaction arises from magnetic compressions in turbulent media. 
{\xu This is the common feature of 
TTD and mirroring. 
Nevertheless, they are two different processes and should be distinguished.} 


The resonant TTD interaction happens when 
{\xu the parallel phase speed of compressible waves matches the parallel speed of the particle.
Such interactions can induce the second order Fermi acceleration of $p_\|$. 
The resulting stochastic increase of $\mu$
associated with TTD acceleration should be distinguished from the pitch-angle diffusion due to resonant scattering.
In contrast, the mirror diffusion
does not require any resonance condition.
It can take place even in the limit case of stationary magnetic bottles, i.e. $V_A\rightarrow 0$, where stochastic acceleration is not involved. 
As another difference, mirroring can not happen at small pitch angles in the loss cone, 
but there is no such constraint for TTD. 

In brief, TTD is a stochastic acceleration process with acceleration-induced diffusion in $\mu$. 
Mirror diffusion is a diffusion process in space, which can be accompanied by stochastic acceleration. }}


\subsection{\xu Damping effect}

Fast modes can be subject to {\xu ion-neutral collisional damping in partially ionized interstellar phases 
\citep{XLY14,Xuc16,XLb18}
and collisionless
damping in galactic halo.} 
The effect of damping of fast modes on scattering of CRs has been addressed in earlier studies, e.g., 
\citet{YL02,YL04,Xuc16,XLb18}. 
{\xu The effect of damping on mirroring by fast modes will be addressed in our future work. 
In the case when fast modes are severely damped, slow modes can dominate the bouncing of CRs 
(see Sections \ref{sec:incomp} and \ref{ssec: comvarf}).}

\section{Summary}
 

{\xu To achieve a more comprehensive description of CR propagation, 
different diffusion mechanisms depending on the properties of MHD turbulence should be taken into account. 
In this work we identify the mirror diffusion as an essential process of CR propagation. }

The perpendicular superdiffusion of CRs 
originates from the superdiffusion of
magnetic fields induced by Alfv\'{e}nic turbulence. 
This effect makes the trapping of CR within 
magnetic bottles improbable. 
Instead, 
CRs bounce with different magnetic mirrors and 
move diffusively parallel to the turbulent magnetic field. As a result,
both {\xu magnetic mirroring} and pitch-angle scattering contribute to the parallel diffusion. 
The former governs the parallel diffusion of CRs at large pitch angles with $\mu<\mu_c$, 
and the latter is dominant at $\mu>\mu_c$, 
where $\mu_c$ corresponds to the balance between {\xu mirroring} and scattering. 
The two diffusion processes interact with each other through exchanging CRs via the pitch angle diffusion.

In the case when fast modes dominate both scattering and {\xu mirroring}, 
$\mu_c$ increases with $E_\text{CR}$. 
As a result, the diffusion coefficient for scattering by fast modes has a weaker dependence on $E_\text{CR}$ compared to the case without {\xu mirroring}, i.e., $\mu_c = 0$. 
It is not an exact power law function of $E_\text{CR}$,
but can be approximated by $\propto E_\text{CR}^{1/3}$.
The scaling $\propto E_\text{CR}^{1/3}$ is expected 
for isotropic turbulence with the Kolmogorov spectrum. 
However, the scattering by Alfv\'{e}n and slow modes with the Kolmogorov spectrum
is inefficient and the resulting diffusion coefficient depends on $E_\text{CR}$ as $\propto E_\text{CR}^{-3/2}$ due to the scale-dependent turbulence anisotropy. 
Our finding provides the physical justification for the commonly used energy scaling of diffusion coefficient.

With the {\xu mirroring} of CRs taken into account, 
the $90^\circ$ problem of quasilinear
gyroresonant scattering can be solved. 
Moreover,
as the mean free path of bouncing CRs is determined by the size of compressive magnetic fluctuations, 
which cannot exceed the driving scale of turbulence,
the corresponding diffusion is slow.
The energy scaling of the diffusion coefficient of bouncing CRs depends on the anisotropy of pitch angle distribution and the energy dependence of $\mu_c$.

In the vicinity of a CR source, 
the injected CRs can have an anisotropic pitch angle distribution. 
Note that the damping of turbulence can also affect the anisotropic distribution. 
If there are more particles at larger pitch angles, 
the diffusion of bouncing CRs can be further suppressed.

For the galactic environment with a nearly uniform and isotropic distribution of CRs, 
the average diffusion coefficient of bouncing and non-bouncing CRs 
{\xu on scales much larger than their mean free paths} 
is usually determined by the diffusion coefficient of the latter, 
as it is larger than that of bouncing CRs. 
Its energy scaling can be non-universal, 
depending on the properties of interstellar turbulence.


\acknowledgments
S.X. acknowledges the support for 
this work provided by NASA through the NASA Hubble Fellowship grant \# HST-HF2-51473.001-A awarded by the Space Telescope Science Institute, which is operated by the Association of Universities for Research in Astronomy, Incorporated, under NASA contract NAS5-26555. 
\software{MATLAB \citep{MATLAB:2018}}

\appendix
\section{Basic scalings of MHD turbulence}
\label{sec:basic}

\subsection{Scalings of incompressible MHD turbulence}

MHD turbulence is a major agent determining the dynamics of CRs.
The turbulence in magnetized media in typical astrophysical settings is injected at the scale $L$ with the injection velocity $V_L$, and the turbulent energy then cascades down to smaller scales. 
If the injection happens with $V_L>V_A$, where $V_A$ is the Alfv\'{e}n speed, 
this is the case of super-Alfv\'{e}nic turbulence. If $V_L<V_A$, the turbulence is 
sub-Alfv\'{e}nic.
The ratio $M_A=V_L/V_A$ is the Alfv\'{e}n Mach number. $M_A=1$ corresponds to the trans-Alfv\'{e}nic turbulence.

Note that we consider turbulence as a result of energy cascade. 
Therefore the Alfv\'{e}n waves excited by CR 
instabilities, e.g. streaming instabilities (see \citealt{FG04}), 
gyro-resonance instabilities 
(see \citealt{LB06}), 
are not classified as turbulence.

The theory of trans-Alfv\'{e}nic turbulence was developed by 
\citet{GS95}
(henceforth GS95)
in the global system of reference with respect to the mean magnetic field. 
\citet{LV99}
(henceforth LV99) 
presented an alternative way to derive the GS95
theory based on the 
theory of turbulent reconnection of magnetic fields.
More importantly, they found that 
the GS95 theory is only valid in the 
``local system of reference" with respect to the local mean magnetic field averaged over the length scale of interest, 
as confirmed by numerical simulations 
\citep{CV00,MG01,CLV_incomp}.
According to the LV99 theory, 
magnetic reconnection 
takes place within one eddy turnover time, 
which allows the 
turbulent motions to mix up magnetic fields without bending them. 
The unconstrained eddy rotation occurs in the system of reference of the local magnetic field averaged over the eddy scale,
and the eddy is aligned with the local magnetic field. 
The direction of the local magnetic field averaged over a small scale can 
differ significantly from the global mean magnetic field direction averaged over a large scale. 

The turbulent cascade in the direction perpendicular to the local magnetic field remains
Kolmogorov-like with 
\begin{equation}
v_l\approx V_A \left(\frac{l_\bot}{L}\right)^{1/3},
\label{GS95perp}
\end{equation}
where it is taken into account that $V_L = V_A$ for trans-Alfv\'{e}nic turbulence, 
$v_l$ is the turbulent velocity at $l_\perp$,
and $l_\perp$ is the perpendicular size of a turbulent eddy. 
The eddy rotation perpendicular to the local magnetic field induces Alfv\'{e}nic perturbation that propagates along the magnetic field with the speed $V_A$. 
The timescale of this perturbation $l_\|/V_A$ should be equal to the eddy turnover time $l_\perp/v_l$. The corresponding relation
between the parallel and perpendicular sizes of the eddy
\begin{equation}
\frac{l_\|}{V_A}\approx \frac{l_\perp}{v_l},
\label{critbal}
\end{equation}
was termed as the ``critical balance" in GS95.
By combining Eq. \eqref{GS95perp} and Eq. \eqref{critbal}, one can obtain the scale-dependent anisotropy of trans-Alfv\'{e}nic turbulence 
\begin{equation}
l_{\|}\approx L \left(\frac{l_\perp}{L}\right)^{2/3}.
\label{ll}
\end{equation}
Smaller eddies are more elongated along the local magnetic field. 
Note that Eqs.  \eqref{GS95perp} and \eqref{ll} should be understood in the statistical sense. They represent the scaling relations between the most probable values of the quantities involved. 
For instance, on the basis of numerical simulations, 
\citet{CLV_incomp}
provided an analytical fit for the detailed distribution function describing the probability of finding a $l_\perp$ at a given $l_\|$.
This was used later in
\citet{YL02,YL04}
and subsequent studies on CR scattering.

For super-Alfv\'{e}nic turbulence injected with $M_A>1$, 
the magnetic field is of marginal importance at $L$.
The super-Alfv\'{e}nic turbulence is initially  hydrodynamic-like with the isotropic Kolmogorov energy spectrum. 
With the decrease of turbulent velocity along the energy cascade, 
$v_l\sim V_L (l/L)^{1/3}$, 
where $v_l$ is the turbulent velocity at the length scale $l$,
the effect of magnetic field becomes more and more manifested. 
Eventually, at the scale
\citep{Lazarian06}
\begin{equation}
l_A=L M_A^{-3},
\label{eq:A12}
\end{equation}
$v_l$ becomes equal to $V_A$, 
and the turbulence becomes fully magnetohydrodynamic. 
To describe the MHD cascade on scales less than $l_A$ in super-Alfv\'{e}nic turbulence,
$L$ in Eqs. \eqref{GS95perp} and \eqref{ll}
should be replaced by $l_A$. 

For sub-Alfv\'{e}nic turbulence with $M_A<1$,
it was shown in LV99 that at $L$ the turbulence is {\it weak}, and the parallel scale of wave packets remains unchanged, i.e. $l_\|=L$. 
The scaling obtained in LV99 for the weak turbulence under the assumption of 
the isotropic turbulence driven at $L$ is 
\begin{equation}
v_l\approx V_L \left(\frac{l_\perp}{L}\right)^{1/2},
\end{equation}
and this result was supported by the subsequent study by 
\citet{Gal00}.
With the decrease of $l_\perp$, 
the intensity of interactions of Alfv\'{e}nic perturbations increases.
At a scale (LV99)
\begin{equation}
l_\text{tran}\approx L M_A^2,
\label{ltrans}
\end{equation} 
where $M_A<1$, the turbulence gets strong. 
For the sub-Alfv\'{e}nic MHD turbulence at $l<l_\text{tran}$,
LV99 derived the relations
\begin{equation}
v_l\approx V_L \left(\frac{l_\bot}{L}\right)^{1/3} M_A^{1/3},
\label{vAlf}
\end{equation}
and 
\begin{equation}
l_{\|}\approx L \left(\frac{l_\bot}{L}\right)^{2/3} M_A^{-4/3},
\label{lpar}
\end{equation}
which at $M_A=1$ can recover the relations in 
Eqs. \eqref{GS95perp} and \eqref{ll}
for trans-Alfv\'{e}nic turbulence.

To generalize our analysis for diffusion of bouncing particles in trans-Alfv\'{e}nic turbulence to super- and sub-Alfv\'{e}nic turbulence, 
we introduce the following effective injections scales. 

(1) Super-Alfv\'{e}nic turbulence.
As discussed above, the super-Alfv\'{e}nic turbulence at $l<l_A$ is equivalent to the trans-Alfv\'{e}nic turbulence injected at $l_A$. 
So by using $l_A$ as the effective injection scale to replace $L$ in trans-Alfv\'{e}nic turbulence,
the results derived for trans-Alfv\'{e}nic are still applicable to super-Alfv\'{e}nic on scales less than $l_A$.


(2) Sub-Alfv\'{e}nic turbulence.
By introducing an effective injection scale 
\begin{equation}
    L_\text{eff}=L M_A^{-4},
    \label{eq:A11}
\end{equation}
Eqs. \eqref{vAlf} and \eqref{lpar} can be rewritten as
\begin{equation}
v_l\approx V_A \left(\frac{l_\bot}{L_\text{eff}}\right)^{1/3},
\label{vAlf1}
\end{equation}
and
\begin{equation}
l_{\|}\approx L_\text{eff} \left(\frac{l_\bot}{L_\text{eff}}\right)^{2/3},
\label{lpar1}
\end{equation}
which take the same forms as Eqs. \eqref{GS95perp} and \eqref{ll}
with $L$ replaced by $L_\text{eff}$.
Therefore, by using $L_\text{eff}$ instead of $L$,
the expressions for trans-Alfv\'{e}nic turbulence can be applied to sub-Alfv\'{e}nic turbulence on scales less than $l_\text{tran}$.

\subsection{Magnetic compressions in MHD turbulence}

The magnetic mirroring effect arise from the magnetic compressions in MHD turbulence, which are associated with slow and fast modes. 
We use the term ``modes" rather than ``waves",
as the nonlinear interactions intrinsic to the turbulent cascade make the properties of magnetic fluctuations different from the properties of linear waves. 
We stress that the compression of magnetic field can also occur in incompressible MHD turbulence. The incompressible limit of slow modes, i.e., pseudo-Alfv\'{e}n modes (see GS95), is a typical example for turbulent compression of magnetic field in an
incompressible medium.

It was pointed out in GS95 that Alfv\'{e}n modes impose their scaling on the pseudo-Alfv\'{e}n modes. With a finite compressibility, slow modes also follow the same scaling as Alfv\'{e}n modes. 
This property was discussed in 
\citet{LG01}
and confirmed numerically by
\citet{CL02_PRL,CL03}.
As discussed above, the anisotropic scaling of slow modes is defined and should be measured in the local system of reference. 
For bouncing CRs that interact with the magnetic compression generated by slow modes, 
they only feel the local magnetic field averaged over the scale of the magnetic compression. 
When describing the gyroresonant scattering of non-bouncing CRs, they also only feel the local magnetic field averaged over the scale comparable to $r_L$. 
The QLT for scattering is 
formally only applicable to infinitesimally small magnetic perturbations.
In the local system of reference, 
CRs only interact with the small magnetic fluctuations on small scales, while the large fluctuations on large scales are not involved. 
This extends the applicability of the QLT.

Fast modes are different from Alfv\'{e}n and slow modes. 
Their evolution is not related to the local system of reference. Therefore, the traditional wave representation is applicable to fast modes. 
The scaling of fast modes is somewhat less certain. 
The theoretical considerations in LG01 for high-$\beta$ plasma and in 
\citet{CL02_PRL,CL03} 
for low-$\beta$ plasma suggested the ``isotropic" energy spectrum of fast modes with $k^{-3/2}$ similar to the acoustic-type turbulence.
However, the simulations in 
\citet{KowL10} 
indicated a steeper spectrum close to $k^{-2}$ for fast modes. 
This discrepancy may be 
attributed to the effect of shocks in the latter study.

\bibliographystyle{aasjournal}
\bibliography{xu}

\end{document}